\documentclass[a4paper,fleqn,usenatbib,useAMS]{mnras}

\usepackage{graphicx}	
\usepackage{amsmath}	
\usepackage{amssymb}	
\usepackage{multicol}        
\usepackage{bm}		
\usepackage{pdflscape}	

\usepackage{booktabs} 

\usepackage[normalem]{ulem}

\usepackage{multirow, array} 

\useunder{\uline}{\ul}{}

\usepackage{float} 

\usepackage[dvipsnames]{xcolor}
\usepackage{tikz,hyperref}

\definecolor{lime}{HTML}{A6CE39}
\DeclareRobustCommand{\orcidicon}{
	\begin{tikzpicture}
	\draw[lime, fill=lime] (0,0) 
	circle [radius=0.13] 
	node[white] {{\fontfamily{qag}\selectfont \tiny ID}};
	\draw[white, fill=white] (-0.0625,0.095) 
	circle [radius=0.007];
	\end{tikzpicture}
	\hspace{-2mm}
}

\foreach \x in {A, ..., Z}{\expandafter\xdef\csname orcid\x\endcsname{\noexpand\href{https://orcid.org/\csname orcidauthor\x\endcsname}
			{\noexpand\orcidicon}}
}

\usepackage[T1]{fontenc}
\usepackage{ae,aecompl}

\usepackage{newtxtext,newtxmath}

\title[Peering into the SySt CH~Cyg]{Peering into the central region of a nano-quasar: {\it XMM-Newton} and {\it Chandra} views of the CH\,Cyg Symbiotic System}

\author[J. A. Toal\'{a} et al.]{J.~A.\,Toal\'{a}$^{1\orcidB}$\thanks{E-mail:\,j.toala@irya.unam.mx}, O.~Gonz\'{a}lez-Mart\'{i}n$^{1\orcidG}$, M.~Karovska$^{2\orcidD}$, R.\,Montez~Jr.$^{2\orcidE}$, M.~K.~Botello$^{3\orcidA}$ and  L.~Sabin$^{3\orcidC}$ \\
$^{1}$Instituto de Radioastronom\'{i}a y Astrof\'{i}sica, UNAM, Antigua Carretera a P\'{a}tzcuaro 8701, Ex-Hda. San Jos\'{e} de la Huerta, Morelia 58089, Mich., Mexico\\
$^{2}$Center for Astrophysics | Harvard \& Smithsonian, 60 Garden Street, Cambridge, MA 02138, USA\\
$^{3}$Instituto de Astronom\'{i}a, Universidad Nacional Aut\'{o}noma de M\'{e}xico, Ensenada 22860, Baja California, Mexico
}

\pubyear{2023}

\begin{document}
\label{firstpage}
\pagerange{\pageref{firstpage}--\pageref{lastpage}}
\maketitle

\begin{abstract}
We present the analysis of archival {\it XMM-Newton} and {\it Chandra} observations of CH\,Cyg, one of the most studied symbiotic stars (SySts). The combination of the high-resolution {\it XMM-Newton} RGS and {\it Chandra} HETG X-ray spectra allowed us to obtain reliable estimates of the chemical abundances and to corroborate the presence of multi-temperature X-ray-emitting gas. Spectral fitting of the medium-resolution {\it XMM-Newton} MOS (MOS1+MOS2) spectrum required the use of an additional component not seen in previous studies in order to fit the 2.0--4.0~keV energy range. Detailed spectral modelling of the {\it XMM-Newton} MOS data suggests the presence of a reflection component, very similar to that found in active galactic nuclei. The reflection component is very likely produced by an ionised disk (the accretion disk around the white dwarf) and naturally explains the presence of the fluorescent Fe emission line at 6.4~keV while also contributing to the soft and medium energy ranges. The variability of the global X-ray properties of CH\,Cyg are discussed as well as the variation of the three Fe lines around the 6--7~keV energy range. We conclude that reflection components are needed to model the hard X-ray emission and may be present in most $\beta/\delta$-type SySt.
\end{abstract}


\section{INTRODUCTION}
\label{sec:intro}

Symbiotic stars (SySts) are binary systems consisting of a red giant and an accreting compact object which could be either a white dwarf (WD) or a neutron star \citep[e.g.,][]{Luna2013,Enoto2014}. WD symbiotics are particularly interesting because accretion might increase the mass of the WD to the Chandrasekhar limit making them potential type Ia supernova progenitors. It is mostly accepted that the accretion process in SySts takes place through Bondi-Hoyle accretion, but the formation of an accretion disk is expected and has been indirectly observed \citep{Livio1984,Alexander2011}. In addition, if the material accumulates enough angular momentum jets will be launched \citep[see][for discussion of the {\it jet feedback mechanism}]{Soker2016}.

X-ray emission has been used to characterise the physics behind the complex circumstellar medium around WD symbiotics. 
The production of X-ray-emitting plasma can be attributed to different origins \citep[see][and references therein]{Mukai2017}. Super soft X-ray emission ($E<$0.4~keV) can be produced by nuclear burning at the surface of the WD \citep{Orio2007} or accreted material hitting its surface \citep{Aizu1973}. Other mechanisms include an accretion disk-like corona \citep{Wheatley2005}, shocks between the WD's wind and that of the red giant companion \citep{Kenny2005} or as a result of the presence of a reflection component \citep{Ishida2009}. However, we note that the details of the spectral properties of the X-ray emission also depend on the stellar masses and accretion rates \citep{Kylafis1982}.

Using {\it ROSAT} observations \citet{Muerset1997} presented the first X-ray spectral classification scheme of SySts, which was  latter extended by the work of \citet{Luna2013} using {\it Swift} data. They defined $\alpha$-type X-ray-emitting SySts as those with spectra dominated at $E<0.4$~keV, 
$\beta$-type was assigned to those SySts with X-ray spectra peaking at 0.8~keV and attributed to the wind collision region, a $\gamma$-type was defined for those SySts with harder X-ray spectra. Finally, $\delta$ classification is used to denote those SySts with hard X-ray emission from the innermost accretion region. A summary of the classification of X-ray detected SySts has been presented in \citet{Merc2019}.

\begin{figure*}
\begin{center}
  \includegraphics[angle=0,width=\linewidth]{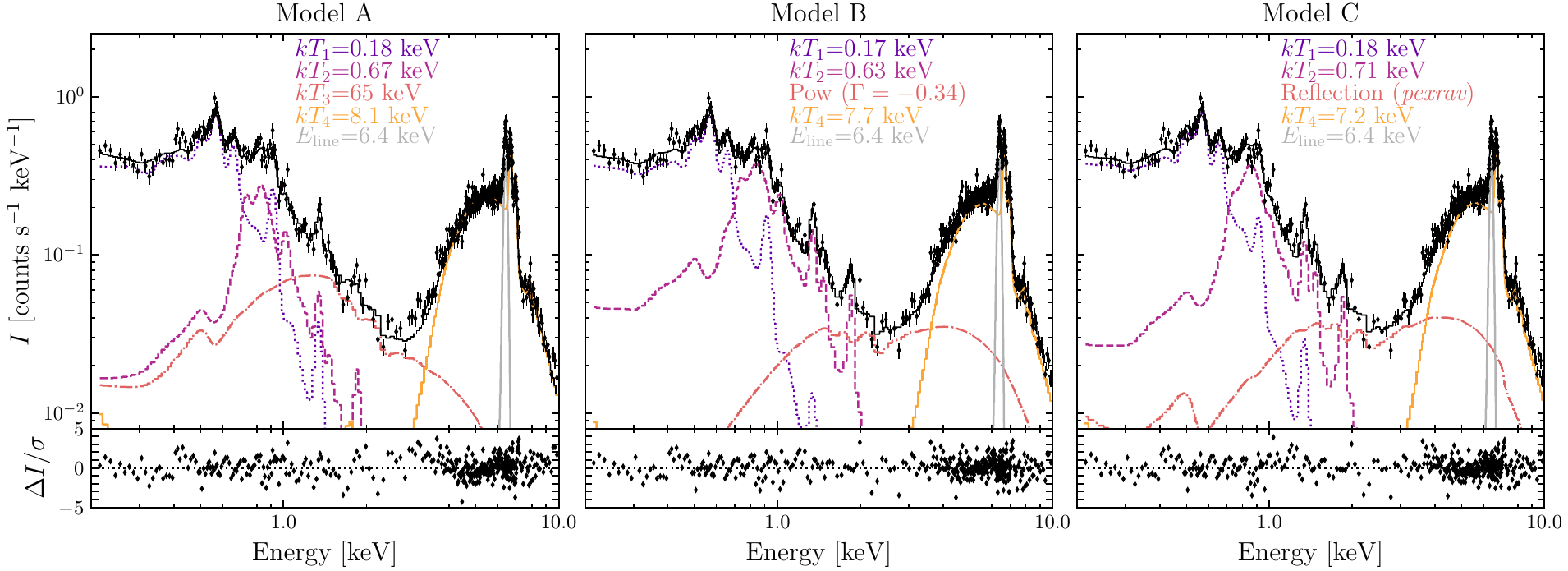}
\end{center}
\caption{{\it XMM-Newton} MOS (MOS1+MOS2)
  spectrum of CH\,Cyg. The black diamonds correspond to the observed spectrum while the black histogram represents the best-fit model with the bottom panels showing the residuals. The different components are illustrated with different (coloured) lines. Details of the models can be found in Section~\ref{sec:epic-spec} and Table~\ref{tab:EPIC_par}.}
\label{fig:spec}
\end{figure*}

One of the X-ray-emitting SySt that has gained much attention is CH\,Cyg. Its high variability is very likely linked to the irregular mass ejections \citep[see, e.g.,][]{Iijima2019,Tarasova2021}. It is located at a relatively close distance from the Sun \citep[$d$=200$\pm$5~pc;][]{BailerJones2021} and as a consequence its mass ejection and jet formation have been resolved in radio, optical and X-ray data \citep{Corradi2001,Solf1987,Taylor1986,Galloway2004,Karovska2007}.

CH\,Cyg has been observed by multiple X-ray missions, including {\it EXOSAT}, {\it ASCA}, {\it   Chandra} and {\it Suzaku}, through different epochs \citep[see][]{Leahy1987,Leahy1995,Ezuka1998,Karovska2007,Mukai2007}. \citet{Muerset1997} classified it as a $\beta$-type source but noted that a detailed spectral fitting was difficult to be achieved \citep[see also ][]{Leahy1995}. Later, \citet{Ezuka1998} showed that its {\it ASCA} spectrum rendered a more complex source. Thus a classification of $\beta/\delta$ seemed to be more appropriate.

\citet{Mukai2007} compared the {\it Suzaku} data with a model that  included two negligibly-absorbed thermal components ($kT_1\sim0.2$~keV, $kT_2\sim0.7$~keV) that roughly fit the spectral range below 2.0~keV. A third, highly-absorbed hotter component ($kT_3\sim$10~keV) was used to fit the more energetic part of the spectra. We note that these authors warned that this might not be the best model to the data. 
In addition, \citet{Mukai2007} combined the available X-ray data at the time and found that the hard X-ray emission was declining. These authors suggested that the decline of the fluorescent Fe line is due to changes in the accretion disk (thickness or precession).

Jet activity was reported in CH\,Cyg using {\it Chandra} data \citep{Galloway2004,Karovska2007} and follow-up {\it Chandra}, {\it HST} and VLA observations presented in \citet{Karovska2010} resolved a jet and its counter part, with angular distances to the central source of 3 and 1~arcsec, respectively. The high spatial resolution {\it Chandra} observations allowed for spectral characterisation of the jets, which mainly emit soft X-rays in the 0.2--2.0~keV energy range. The central unresolved source has an angular size of 0.5~arcsec in radius and its spectrum was typical of the $\beta/\delta$-type SySt \citep[see][]{Karovska2010}.

In this paper we analyse publicly available {\it XMM-Newton} and {\it Chandra} observations of CH\,Cyg. We present the first high-resolution X-ray spectra of CH\,Cyg. The data are analysed similarly to previous X-ray studies of CH\,Cyg. We propose that some of the spectral properties can be explained by the combination of an accretion disk, a reflection component and soft extended emission, very similar to what is currently seen in active galactive nuclei (AGN) and suggested for other SySts.

This paper is organized as follows. In Section~\ref{sec:observations} we present the observations and their preparation. Section~\ref{sec:results} presents the results of our spectral modelling. A discussion addressing the general properties of the X-ray emission from CH\,Cyg, its associated variability and comparison with other sources is presented in Section~\ref{sec:diss}. Our final comments and conclusions are presented in Sec.~\ref{sec:conclusions}.

\section{Observations and data preparation}
\label{sec:observations}

\subsection{{\it XMM-Newton} data}

CH\,Cyg was observed by {\it XMM-Newton} on 2018 May 24 (Ob.\,ID.:
0830190801; PI: N. Schartel) for a total observing time of
36.1~ks. The observations were obtained using the thick optical
blocking filter on the small window mode. Data were obtained using the European
Photon Imaging Cameras (EPICs) and the reflection grating spectrometers
(RGSs). The total exposure time for the EPIC-pn, MOS1 and MOS2 cameras was 34.2~ks, 34.7 and 34.7~ks, respectively. The RGS1 and RGS1 were used for 35.0~ks each. The data presented in this work were processed with the Science Analysis Software \citep[{\sc sas};][]{Gabriel2004} 
version 20.0\footnote{\url{https://www.cosmos.esa.int/web/xmm-newton/sas-news}} with the calibration files obtained on 2022 August 16.

The event files of the EPIC cameras were created using the {\sc sas} tasks {\it emproc} and {\it epproc}. In order to evaluate periods of high background signal, we extracted light curves in the 10--12~keV energy range and examined the background levels. We rejected time intervals where the background count rate
was higher than 0.40, 0.10 and 0.15~counts s$^{-1}$ for the pn, MOS1 and MOS2 cameras. This process reduced the useful exposures to 24.0, 33.3 and 33.0~ks, for the pn, MOS1 and MOS2 cameras, respectively.

We extracted medium-resolution spectra from the three EPIC cameras taking a circular aperture with a radius of 20~arcsec centered on CH\,Cyg. The background was extracted from a region with the same size and no contribution from other point sources in the vicinity of this SySt. The resultant background-subtracted count rate was 2750$\pm$12, 655$\pm$5 and 675$\pm$5~counts~ks$^{-1}$ corresponding to the 0.2--10.0~keV energy range for the pn, MOS1 and MOS2 cameras, respectively.

After evaluating the data using the {\sc sas} task {\it epatplot} we determined that the pn exposure was affected by pile-up. The MOS exposures were not affected by pile-up and have sufficient signal-to-noise and quality to perform spectral analysis. We combined the two MOS spectra and their corresponding background and calibration matrices using the {\sc sas} task {\it epicspeccombine}. The background-subtracted MOS (MOS1+MOS2) was binned to a minimum of 80 counts per bin. The resultant MOS spectrum is presented in Fig.~\ref{fig:spec}.

\begin{figure*}
\begin{center}
  \includegraphics[angle=0,width=0.95\linewidth]{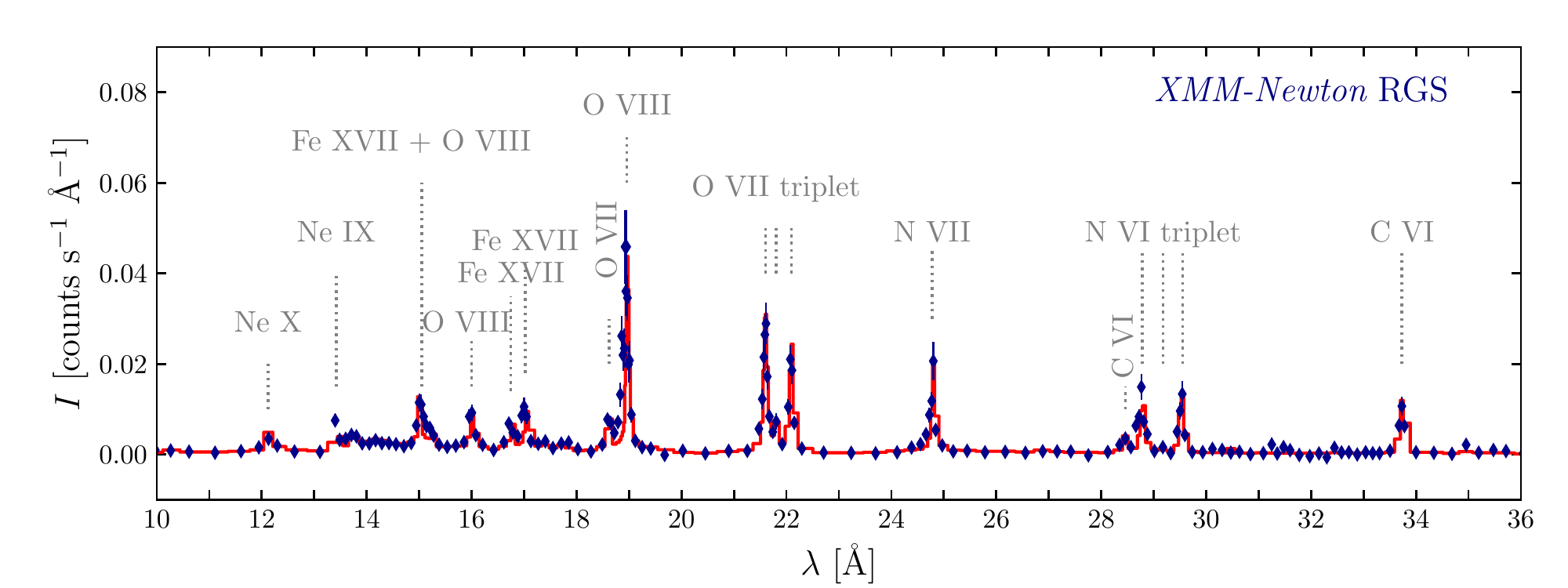}\\
  \includegraphics[angle=0,width=0.95\linewidth]{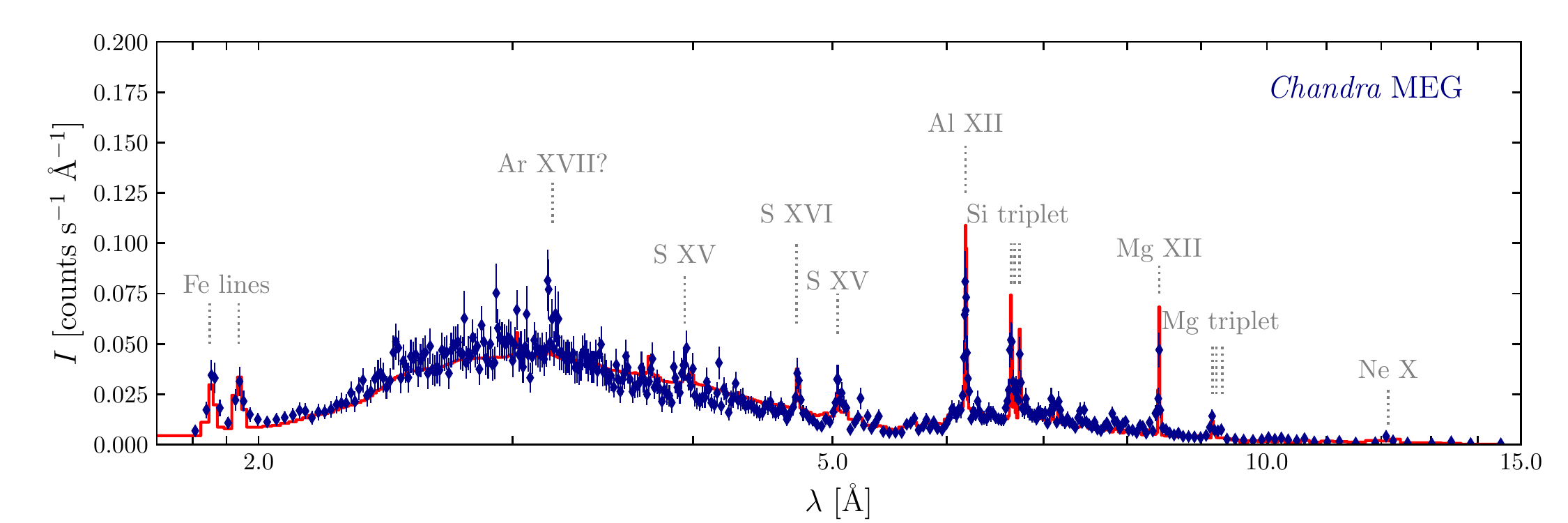}\\
  \includegraphics[angle=0,width=0.95\linewidth]{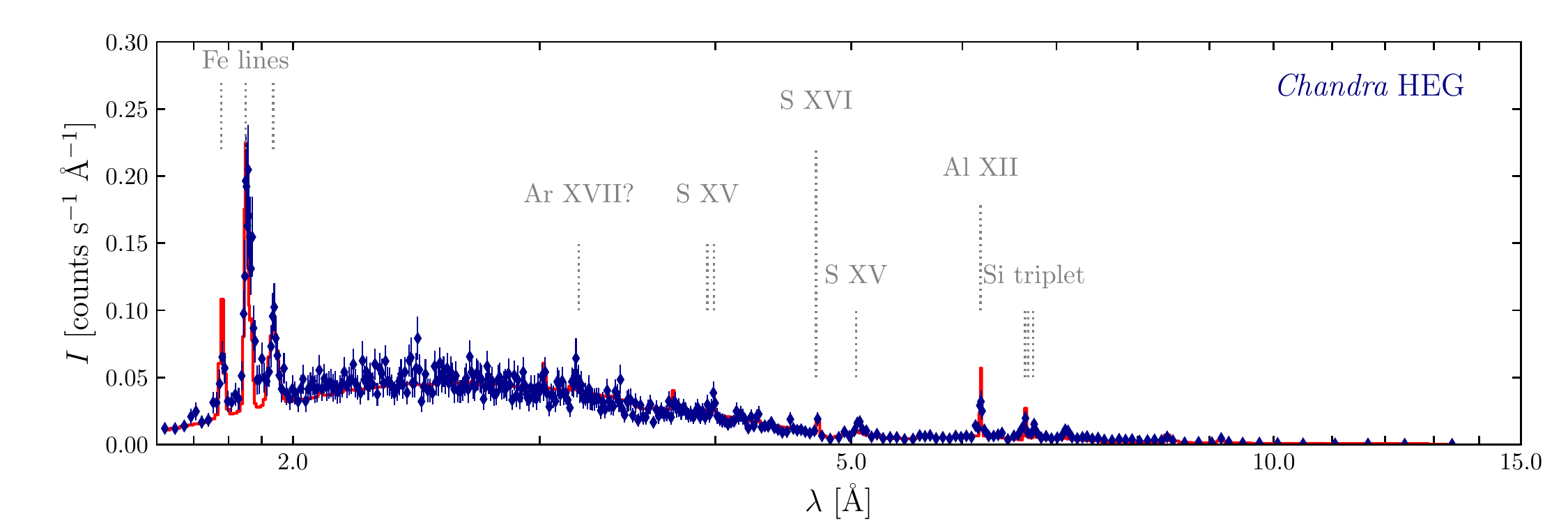}
\end{center}
\caption{High-resolution X-ray spectra of CH\,Cyg (diamonds with error bars). The solid (red) line represents the best fit to the data. Different emission lines are labelled. The solid, dotted and dashed grey lines in the HEG and MEG spectra represent the three components of the models detailed in Table~\ref{tab:RGS_par}.}
\label{fig:rgs_spec}
\end{figure*}

The RGS source and background spectra as well as calibration matrices were produced using the {\sc sas} task {\it rgsproc}\footnote{See \url{https://xmm-tools.cosmos.esa.int/external/xmm_user_support/documentation/sas_usg/USG/rgsproc.html}}. The net count rates for the 5--36~\AA\, wavelength range of the 1st order RGS1 and RGS2 instruments resulted in 54.3$\pm$1.9 and 54.6$\pm$1.7~counts~ks$^{-1}$, respectively. For both 1st order RGS spectra we obtained effective exposure times of 34.9~ks. The signal-to-noise of the 2nd order RGS spectra are poor and will not be considered further. 
The 1st order spectra of the RGS1 and RGS2 instruments were combined to create a single RGS spectrum using the {\sc sas} task {\it rgscombine}. The resultant 1st order RGS(1+2) spectrum was binned to a minimum of 30 counts per bin and its presented in the top panel of Fig.~\ref{fig:rgs_spec}.

In addition, and for means of discussion and comparison, we use the {\it XMM-Newton} observations of the well-known SySt R~Aqr. These {\it XMM-Newton} observations of R~Aqr were obtained on 2005 June 30 (PI: E. Kellogg; Obs.\,ID.: 0304050101) and consist only of EPIC observations. A preliminary analysis of these observations was presented in \citet{Toala2022} and we point the reader to that paper for details. For the present paper, we extracted pn, MOS1 and MOS2 spectra from circular apertures with radii of 1.5~arcmin centred on R\,Aqr. These extraction regions include the contribution of the most extended soft emission, the jets and the central source \citep[see][]{Kellogg2007,Nichols2007,Toala2022}. A final EPIC (pn+MOS1+MOS2) spectrum was created and its shown in Fig.~\ref{fig:spec_r_aqr} of Appendix~\ref{sec:app}.

\subsection{{\it Chandra} high-resolution data}

CH\,Cyg has been observed on several occasions by different instruments on the {\it Chandra X-ray Observatory}. In this paper we present the analysis of the High Energy Transmission Gratings (HETG). Data collected using the Medium and High Energy Gratings (HEG and MEG) of CH\,Cyg were obtained on 2009 Nov 25 and Dec 1--3. These correspond to Obs.\,IDs. 9969, 12032, 12034, and 12035 (PI: K. Mukai). The HETG observations amount to a total exposure time of 100~ks.

The reduced data were retrieved from the Transmission Grating Catalog and Archive \citep[TGCat;][]{Huenemoerder2011}\footnote{\url{http://tgcat.mit.edu/}.}. The TGCat provides source spectra as well as calibration matrices for each observation. The different orders from all observations were combined using the {\sc ciao} task {\it combine\_grating\_spectra}\footnote{See details at \url{https://cxc.cfa.harvard.edu/ciao/ahelp/combine_grating_spectra.html}.}. The resultant {\it Chandra} MEG and HEG spectra are shown in the middle and bottom panels of Fig.~\ref{fig:rgs_spec}, respectively.

\begin{table}
\begin{center}
\setlength{\columnwidth}{0.1\columnwidth}
\setlength{\tabcolsep}{0.4\tabcolsep}
\caption{Details of the best-fit models of the MOS spectrum of CH\,Cyg compared to the best-fit model obtained for R~Aqr. The total (0.2--10.0~keV) and soft (0.2--2.2~keV) observed ($f_\mathrm{X}$) and unabsorbed ($F_\mathrm{X}$) fluxes are in cgs units (erg~s$^{-1}$~cm$^{-2}$).}
\label{tab:EPIC_par}
\begin{tabular}{lcccc}
\hline
                                      & CH Cyg             & CH Cyg            & CH Cyg            & R Aqr              \\
                                      & MOS                & MOS               & MOS               & EPIC \\
                                      & Model A            & Model B           & Model C           &                    \\
Parameter                             & 0.2--10.0~keV      & 0.2--10.0~keV   & 0.2--10.0~keV   & 0.2--10.0~keV    \\  
\hline
$\chi^2$                              & 1.75               & 1.51              &  1.57             & 0.68               \\
$N_\mathrm{H,1}$[10$^{22}$~cm$^{-2}$] & $<$0.01            & 0.13$\pm$0.01     &  0.03$\pm$0.01    & 0.003$\pm$0.02     \\
$kT_1$          [keV]                 & 0.18$\pm$0.01      & 0.18$\pm$0.01     & 0.18$\pm$0.01     & 0.14$\pm$0.01      \\
$A_1$           [cm$^{-5}$]           & 1.9$\times10^{-3}$ & 2.4$\times10^{-3}$& 8.3$\times10^{-4}$ & 1.1$\times10^{-4}$ \\
$kT_2$          [keV]                 & 0.67$\pm$0.02      & 0.63$\pm$0.01     & 0.71$\pm$0.01      & 0.65$\pm$0.11      \\
$A_2$           [cm$^{-5}$]           & 3.2$\times10^{-4}$ & 7.5$\times10^{-4}$& 5.1$\times10^{-4}$ & 6.0$\times10^{-6}$ \\
$kT_3$          [keV]                 &  65$\pm$100        & \dots             &  \dots             & \dots              \\
$\Gamma$                              & \dots              & $-$0.34$\pm$0.12  & 1.9$\pm$0.2        & $-$0.44$\pm$0.40   \\
$\mathrm{Abund_{ref}}$                & \dots              & \dots             & $<$0.64        & \dots              \\
$A_3$           [cm$^{-5}$]           & 1.7$\times10^{-3}$ & 6.4$\times10^{-5}$& 2.8$\times10^{-2}$              & 8.2$\times10^{-7}$ \\
$N_\mathrm{H,2}$[10$^{22}$~cm$^{-2}$] & 50$\pm$1.5         & 55.6$\pm$0.8      & 71$\pm$2       & 60$\pm$13          \\
$kT_4$   [keV]                        & 8.1$\pm$0.2        & 7.7$\pm$0.2       & 7.2$\pm$0.2        & 9.4$\pm$1.9        \\
$A_4$    [cm$^{-5}$]                  & 0.21               & 0.21              & 0.26               & 8.4$\times10^{-4}$ \\
$E_\mathrm{line}$ [keV]               & 6.41$\pm$0.01      & 6.42$\pm$0.01     &  6.42$\pm$0.01     & 6.42$\pm$0.02      \\
$\sigma$         [keV]                & 6.5$\times10^{-2}$ & 6.2$\times10^{-2}$& 5.7$\times10^{-2}$ & 5.1$\times10^{-2}$ \\
$A_\mathrm{line}$ [cm$^{-5}$]         & 1.3$\times10^{-3}$ & 1.4$\times10^{-3}$& 5.3$\times10^{-4}$ & 6.9$\times10^{-6}$ \\
\hline
C                                     & 1.0                & 1.0               & $<$0.87        & 1.3$\pm$1.1        \\
N                                     & 2.9$\pm$0.5        & 2.6$\pm$0.5       & 2.3$\pm$0.5        & 2.1$\pm$1.4        \\
O                                     & 0.8$\pm$0.1        & 0.7$\pm$0.1       & 0.7$\pm$0.1        & 0.5$\pm$0.3        \\
Mg                                    & 6.3$\pm$2.2        & 1.5$\pm$0.2       & 2.6$\pm$0.3        & 1.0                \\
Si                                    & 0.5$\pm$0.5        & 1.4$\pm$0.2       & 1.9$\pm$0.3        & 1.0                \\
Fe                                    & 0.9$\pm$0.2        & 0.5$\pm$0.1       & 0.84$\pm$0.05        & 0.3$\pm$0.2        \\
\hline
$f_\mathrm{X,soft}$  & 5.0$\times$10$^{-12}$  & 4.9$\times$10$^{-12}$  & 4.9$\times$10$^{-12}$ & 2.1$\times10^{-13}$\\
$F_\mathrm{X,soft}$  & 1.7$\times$10$^{-10}$  & 1.7$\times$10$^{-10}$  & 2.0$\times$10$^{-10}$ & 9.2$\times10^{-13}$\\
$L_\mathrm{X,soft}$ [erg~s$^{-1}$] & 8.1$\times10^{32}$ & 8.1$\times10^{32}$ & 9.5$\times10^{32}$ & 1.6$\times10^{31}$\\
\hline
$f_\mathrm{X,TOT}$   & 7.5$\times$10$^{-11}$  & 7.7$\times$10$^{-11}$  & 7.6$\times$10$^{-11}$ & 6.2$\times10^{-13}$\\
$F_\mathrm{X,TOT}$   & 4.6$\times$10$^{-10}$  & 4.7$\times$10$^{-10}$  & 5.1$\times$10$^{-10}$ & 2.3$\times10^{-12}$\\
$L_\mathrm{X}$ [erg~s$^{-1}$]      & 2.2$\times10^{33}$ & 2.2$\times10^{33}$ & 2.4$\times10^{33}$ & 4.1$\times10^{31}$\\
\hline
\end{tabular}
\end{center}
\end{table}

\section{Spectral analysis}
\label{sec:results}

In this section we present a number of models to try to assess the general properties of the X-ray emission from CH\,Cyg. Here we want to demonstrate the need of a more complex model than that proposed in \citet{Mukai2007} for CH~Cyg. The models presented here are statistically-acceptable models. However, a physically driven model which explains most of the spectral features of the X-ray spectrum of CH~Cyg is presented in Section~\ref{sec:diss}.

All of the spectral fits were attempted using the X-Ray Spectral Fitting Package \citep[{\sc xspec};][]{Arnaud1996}\footnote{\url{https://heasarc.gsfc.nasa.gov/xanadu/xspec/}}.
Fits to the X-ray emission detected from CH\,Cyg were attempted by using absorbed optically-thin 
emission plasma models. This was done initially by adopting solar abundances \citep{Lodders2009} and the {\it tbabs} absorption model\footnote{\url{https://heasarc.gsfc.nasa.gov/xanadu/xspec/manual/node268.html}}  \citep{Wilms2000}. The optically-thin emission plasma components were modelled using {\it apec} models\footnote{This is an emission spectrum computed from collisionally-ionised diffuse gas. See the AtomDB atomic database at \url{http://atomdb.org/}.}. As is shown below, some models have similar quality fits when adopting a power law component. The goodness of a fit was evaluated by assessing the reduced $\chi^{2}$ statistics.

\subsection{The MOS spectrum}
\label{sec:epic-spec}

Fig.~\ref{fig:spec} presents the medium-resolution MOS spectrum of CH\,Cyg, which resembles previously published X-ray spectra of this SySt (see references listed in Sec.~\ref{sec:intro}): a dominant soft X-ray component below 3.0~keV with evident contribution from spectral lines plus the contribution from harder X-ray emission that exhibits the clear presence of the Fe complex (see Fig.~\ref{fig:spec}).

We started by fitting models with two slightly-absorbed plasma temperatures plus a highly-absorbed plasma component. According to \citet{Mukai2007}, these components are needed to fit the soft and hard energy ranges in the 0.2--2.0 and 4.0--10.0~keV, respectively. Similar models as those suggested by \citet{Mukai2007} did not result in good quality fits. The best models required an extra component to fit the 2.0--4.0~keV energy range. This seems to be the case of $\beta/\delta$-type SySts when good-quality X-ray spectra are detected \citep[see also the case of NQ~Gem discussed in][]{Toala2023}. In addition, the emission from the 6.4~keV fluorescent line was modelled by a Gaussian component.

\begin{figure*}
\begin{center}
  \includegraphics[angle=0,width=\linewidth]{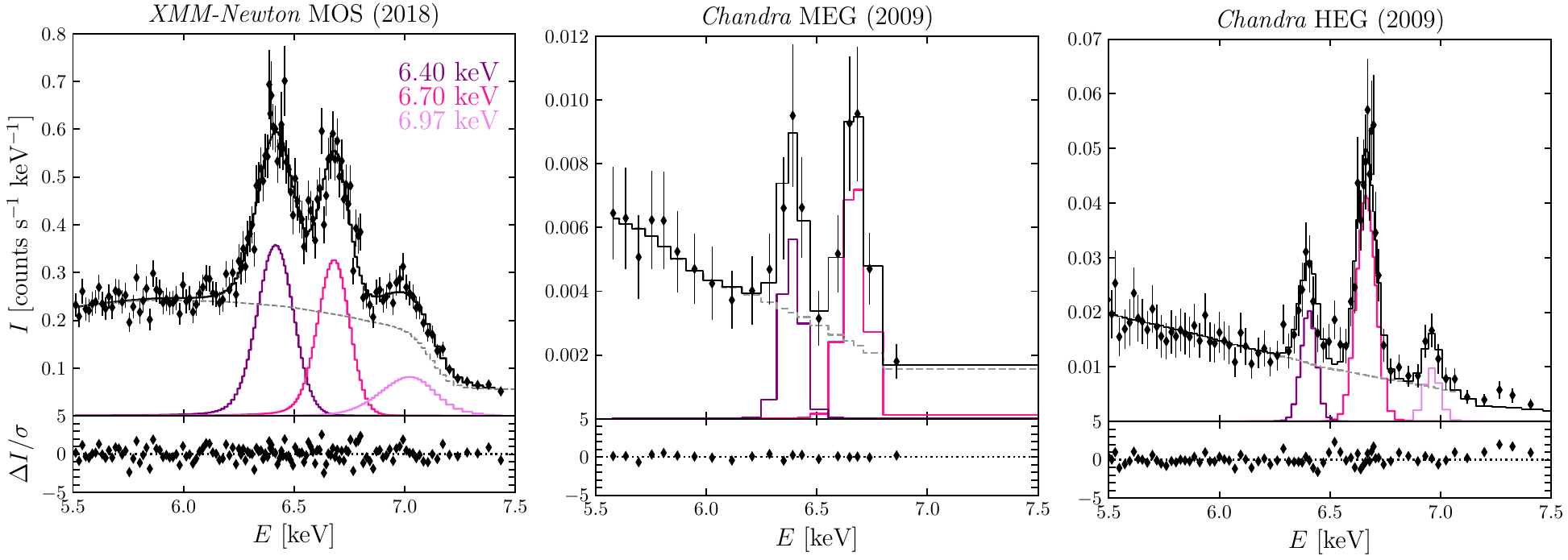}
\end{center}
\caption{X-ray spectra of CH\,Cyg in the 5.5--7.5~keV energy range showing the three Fe emission lines. The black histogram corresponds to the best-fit model and the components} are shown with different {\bf coloured} lines: the grey (dashed) line corresponds to the bremsstrahlung component and purple shades for the three Fe lines. See Section~\ref{sec:Fe_lines} for details.
\label{fig:Fe_lines}
\end{figure*}

A relatively good spectral fit ($\chi^{2}$=1.75; labelled as Model~A) was achieved by adopting the following components:
\begin{equation}
    {\rm tbabs}_1\times ({\rm apec}_1+{\rm apec}_2+{\rm apec}_3) + {\rm tbabs}_2 \times ({\rm apec}_4 + {\rm Gauss}).
\end{equation}
\noindent However, {\sc xspec} was not able to appropriately fit the plasma temperature of the third component (see Table~\ref{tab:EPIC_par}). A better fit was achieved by replacing the third plasma component (${\rm apec}_3$) with a power law (labelled as Model~B). The goodness of fit improved to a reduced $\chi^{2}$=1.51. We note here that a power law component is usually associated with non-thermal emission, which is not expected from SySts \citep[see][]{Mukai2017}. That is, although this component improves the fit, it might not have a physical origin.

Models A and B require the presence of the Fe K$\alpha$ emission line at 6.4 keV which is fitted with a Gaussian. We interpret this emission as a signature of reprocessed emission in the surrounding medium, a well-known process found in the X-ray spectra of accreting systems such as SySts and active galactic nuclei \citep[AGN; see][]{Eze2014}.
Thus, in Model~C, we also test a scenario in which the power-law model is replaced by a reflection component. In particular, we used the {\sc xspec} reflection model {\it pexrav} \citep{Magdziarz1995} from neutral material (but we note that an ionised reflection using {\it pexriv} produces similar results). Model~C produces a statistically similar fit to Model~B ($\chi^{2}$=1.57), but with a photon index of $\rm{\Gamma=2}$. We note that similar models including the presence of a neutral medium that produces a reflection have been tested in the literature. In particular, the works of \citet[][]{Luna2019_reflec} and 
 \citet{Ishida2009} use the {\sc xspec} {\it reflect} model which is a generalised version of the {\it pexrav} component used here.

The parameters of models A, B and C are provided in Table~\ref{tab:EPIC_par} and Fig.~\ref{fig:spec} compares these models with the MOS spectrum of CH\,Cyg in the 0.2--10.0~keV energy range. Observed ($f_\mathrm{X}$) and unabsorbed ($F_\mathrm{X}$) fluxes as well as X-ray luminosities ($L_\mathrm{X}$) calculated for the complete energy range (0.2--10.0~keV) and the soft band (0.2--2.2~keV) are also listed at the bottom lines of Table~\ref{tab:EPIC_par}. The three spectral fits initially included the abundances of C, N, O, Mg, Si and Fe as free parameters and their corresponding values are also listed in this table. Abundances were set to solar when they could not be constrained.

We note that the reflection component of Model C depends on the spectral shape of the emitting source and on the geometry and composition of the medium reprocessing the emission. In particular, the {\it pexrav} component is calculated for a source emitting as an exponentially cut off power law and it assumes that the reflector is a plane-parallel slab perpendicular to the emitting source. In order to test geometrically-motivated models, we also tested the smooth torus-like model component named {\it borus02} \citep{Balokovic2018}. The fit is  worse than Model~C ($\chi^{2}$=1.71), with residuals due to emission lines not shown in the data that might suggest an incorrect assumption of the chemical composition of the torus. Indeed, {\it borus02} model was designed to describe reflection from cold and neutral tori of AGNs. Thus, its use for CH Cyg is not completely justified. Despite that, it is interesting to note that the half opening angle of the torus of this model is restricted to less than 5 degrees, consistent with the reflection happening in a thin disk. A physically-motivated model for CH Cyg is discussed in Section~\ref{sec:model_disk}.

\begin{table}
\setlength{\columnwidth}{0.1\columnwidth}
\setlength{\tabcolsep}{0.6\tabcolsep}
\caption{Fe K$\alpha$ lines detected in the MOS, MEG and HEG spectra of CH\,Cyg. $f$ and $F$ are the observed and unabsorbed fluxes.}
\begin{center}
\begin{tabular}{ccccccc}
\hline
Line   & $E_\mathrm{obs}$  & $\lambda_\mathrm{obs}$& $\sigma$  & $EW$ & $f$                      &  $F$ \\
       &     (keV)          & \AA           & (eV)      & (eV) & (erg~s$^{-1}$~cm$^{-2}$) & (erg~s$^{-1}$~cm$^{-2}$)\\
\hline
MOS  & (2018) \\
6.40    &     6.419  & 1.933 & 47.7  & 226 & (4.8$\pm$0.7)$\times$10$^{-12}$ &  (1.7$\pm$0.2)$\times$10$^{-11}$ \\
6.70    &     6.682  & 1.856 & 24.2  & 186 & (4.5$\pm$0.6)$\times$10$^{-12}$ &  (1.4$\pm$0.2)$\times$10$^{-11}$ \\
6.97    &     7.045  & 1.761 & 125   & 179 & (2.5$\pm$1.0)$\times$10$^{-12}$ &  (8.2$\pm$3.3)$\times$10$^{-12}$ \\
\hline
MEG  & (2009)\\
6.40    &     6.387  & 1.942 & 33.9  & 122 & (1.2$\pm$2.2)$\times$10$^{-12}$ & (3.9$\pm$7.0)$\times$10$^{-12}$\\
6.70    &     6.672  & 1.859 & 36.7  & 312 & (2.8$\pm$4.2)$\times$10$^{-12}$ & (7.6$\pm$11.6)$\times$10$^{-12}$\\
\hline
HEG & (2009)\\
6.40    &     6.411  & 1.935 & 62.5  & 114 & (1.1$\pm$0.1)$\times$10$^{-12}$ & (2.0$\pm$0.4)$\times$10$^{-12}$ \\
6.70    &     6.666  & 1.861 & 39.3  & 376 & (2.9$\pm$0.4)$\times$10$^{-12}$ & (4.7$\pm$0.8)$\times$10$^{-12}$ \\
6.97    &     6.962  & 1.782 & 32.9  & 92  & (7.5$\pm$1.8)$\times$10$^{-13}$ & (1.2$\pm$0.3)$\times$10$^{-13}$ \\ 
\hline
\end{tabular}
\end{center}
\label{tab:Fe_lines}
\end{table}

\subsection{The Fe emission lines}
\label{sec:Fe_lines}

\citet{Mukai2007} gathered all available X-ray observations of CH\,Cyg at the time and demonstrated the variable nature of the three Fe lines. In some cases, the presence of all three lines was only unveiled after careful Gaussian fitting. In particular, the presence of the three Fe lines is difficult to assess in the oldest X-ray observations (1994 {\it ASCA} and 2001 {\it Chandra}), but these are best envisaged in the 2006 {\it Suzaku} spectra \citep[see ][]{Mukai2007,Eze2014}. Still, it seems that the fluorescent Fe line at 6.4~keV has a negligible contribution in the {\it ASCA} and {\it Chandra} data, but became important by 2006.

The MOS spectrum presented here shows the unambiguous presence of the three Fe lines (see Fig.~\ref{fig:Fe_lines} left panel). 
We modelled the 5.5--7.5~keV energy range of the MOS spectrum to calculate the individual line fluxes. Similarly to \citet{Mukai2007}, we modelled the spectrum with an absorbed bremsstrahlung spectrum with the contribution from three Gaussians. The energy of the fluorescent, He-like and H-like Fe lines were initially set to 6.4, 6.7 and 6.97~keV, respectively, but were allowed to vary.

The best fit to this particular energy range resulted in an absorbing column density of $N_\mathrm{H}=50\times10^{22}$~cm$^{-2}$, very similar to that described in the previous section, and the bremsstrahlung model resulted in a temperature of 15.8~keV. The resultant central energy of the line ($E_0$), line width ($\sigma$), equivalent width ($EW$), observed ($f$) and unabsorbed ($F$) fluxes of the three Fe lines are listed in Table~\ref{tab:Fe_lines}. We note that the $EW$ was measured using the {\sc xspec} task {\it eqwidth} before correcting the lines for the absorption.

\begin{table}
\setlength{\columnwidth}{0.1\columnwidth}
\setlength{\tabcolsep}{0.9\tabcolsep}
\caption{Detected emission lines in the high-resolution RGS, MEG and HEG spectra of CH\,Cyg. $f$ corresponds to the observed flux of each line in cgs units of $10^{-14}$~erg~cm$^{-2}$~s$^{-1}$}.
\begin{center}
\begin{tabular}{lcccccc}
\hline
Ion            &$\lambda_\mathrm{lab}$ & $\lambda_\mathrm{obs}$ & $E$ & $\sigma$ & $EW$ & $f$ \\
               & (\AA)  & (\AA)  & (keV) & (eV) & (eV)  & (cgs) \\
\hline
RGS \\
Mg\,{\sc xi}   &  9.21  & 9.22           & 1.345 & 0.01 & 38.5 &1.3$\pm$0.6\\
Ne\,{\sc x}    &  12.13 &  12.13         & 1.021 & 1.2  & 138  & 5.5$\pm$1.1\\
Ne\,{\sc ix}   & 13.45  & 13.43          & 0.923 & 4.5  & 90.9 & 8.9$\pm$1.0\\
Ne\,{\sc ix}   & 13.70  & 13.70          & 0.905 & 4.6  & 27.4 & 3.9$\pm$1.3\\
Fe\,{\sc xviii} & 14.15 & 14.17          & 0.870 & 11.6 & 17.9 & 4.4$\pm$1.8\\
Fe\,{\sc xvii} & 15.02  & 15.02          & 0.823 & 2.5  & 15.1 & 1.8$\pm$0.2\\
O\,{\sc viii}  & 15.17  & 15.20          & 0.820 & 8.5  & 70.4 & 5.5$\pm$2.2\\
O\,{\sc viii}  & 16.00  & 16.00          & 0.776 & 2.0  & 55.5 & 3.4$\pm$0.8\\
Fe\,{\sc xvii} & 16.78  & 16.71          & 0.742 & 1.3  & 11.6 & 1.8$\pm$0.5\\
Fe\,{\sc xvii} & 17.05  & 17.00          & 0.729 & 2.1  & 35.2 & 4.4$\pm$0.7\\
O\,{\sc vii}   & 18.63  & 18.62          & 0.666 & 2.0  & 11.1 & 3.0$\pm$0.6\\
S\,{\sc xvi}   & 18.88  & 18.86          & 0.657 & 0.003& 15.4 & 4.0$\pm$0.8\\
S\,{\sc xv}    & 18.94  & 18.93          & 0.655 & 0.003& 12.2 & 3.3$\pm$1.5\\
O\,{\sc viii}  & 18.96  & 18.96          & 0.654 & 1.1  & 54.2 & 9.6$\pm$1.1\\
O\,{\sc vii }  & 21.60  & 21.61          & 0.574 & 1.1  & 73.8 & 21.6$\pm$1.6\\
O\,{\sc vii }  & 21.80  & 21.81          & 0.568 & 0.003& 5.0  & 3.6$\pm$0.9 \\
O\,{\sc vii }  & 22.10  & 22.09          & 0.561 & 0.5  & 23.7 & 14.5$\pm$1.5\\
N\,{\sc vii}   & 24.78  & 24.80          & 0.500 & 0.5  & 58.7 & 8.2$\pm$0.8 \\

C\,{\sc vi}    & 28.47  & 28.46          & 0.436 & 0.9  & 6.0  & 2.4$\pm$0.7 \\
N\,{\sc vi}    & 28.78  & 28.77          & 0.431 & 1.0  & 38.3 & 11.7$\pm$1.0\\
N\,{\sc vi}    & 29.08  & 29.18          & 0.426 & 0.001& 3.2  & 0.8$\pm$0.5 \\
N\,{\sc vi}    & 29.53  & 29.55          & 0.420 & 0.4  & 21.7 & 6.4$\pm$0.9 \\

C\,{\sc vi}    & 33.73  & 33.73          & 0.368 & 0.5  & 159.3 & 13.0$\pm$2.8\\
\hline
MEG \\
S\,{\sc xv}     & 3.95  &  3.95  & 2.140  & 0.018& 23.6 & 9.6$\pm$3.9 \\
S\,{\sc xvi}    & 4.73  &  4.74  & 2.619  & 7.9  & 31.3 & 10.0$\pm$2.6\\
S\,{\sc vi}     & 5.06  &  5.06  & 2.454  & 16.0 & 80.0 & 18.6$\pm$5.1\\
Al\,{\sc xiii}  & 6.18  &  6.18  & 2.007  & 4.3  & 40.3 & 6.7$\pm$0.7 \\
Mg\,{\sc xii}   & 8.43  &  8.43  & 1.472  & 1.7  & 27.4 & 1.9$\pm$0.3 \\      
Mg\,{\sc xi}    & 9.17  &  9.17  & 1.352  & 4.6  & 21.4 & 1.5$\pm$0.4 \\ 
Mg\,{\sc xi}    & 9.23  &  9.23  & 1.343  & 0.2  & 6.5  & 0.5$\pm$2.0 \\ 
Mg\,{\sc xi}    & 9.31  &  9.31  & 1.331  & 0.3  & 13.3 & 1.0$\pm$0.3 \\ 
\hline
HEG \\
S\,{\sc xv}     & 3.95  &  3.95  & 3.140  & 0.33 & 4.7  & 2.1$\pm$2.0 \\
S\,{\sc xv}     & 3.99  &  4.00  & 3.109  & 6.5  & 17.3 & 7.1$\pm$3.3 \\
S\,{\sc xiv}    & 4.95  &  4.94  & 2.507  & 0.008& 5.8  & 2.4$\pm$1.0 \\
S\,{\sc xv}     & 5.06  &  5.07  & 2.446  & 14.3 & 50.1 & 14.4$\pm$5.0\\
Al\,{\sc xiii}  & 6.18  &  6.18  & 2.006  & 5.1  & 45.8 & 7.1$\pm$0.6 \\
Si\,{\sc xiii}  & 6.65  &  6.64  & 1.870  & 8.6  & 41.8 & 5.9$\pm$2.0 \\
Si\,{\sc xiii}  & 6.69  &  6.69  & 1.853  & 11.4 & 14.9 & 2.8$\pm$2.3 \\
Si\,{\sc xiii}  & 6.74  &  6.75  & 1.839  & 4.2  & 13.3 & 1.8$\pm$2.0 \\
Mg\,{\sc xii}   & 7.11  &  7.11  & 1.746  & 5.7  & 16.3 & 2.0$\pm$1.3 \\

Mg\,{\sc xii}   & 8.42  &  8.39  & 1.477  & 5.5  & 28.3 & 2.0$\pm$1.5\\
Mg\,{\sc xi}    & 9.23  &  9.21  & 1.347  & 1.0  & 33.1 & 2.1$\pm$1.2\\
\hline
\end{tabular}
\end{center}
\label{tab:rgs_lines}
\end{table}

The Fe lines are also detected in the {\it Chandra} MEG and HEG spectra (see middle and right panels in Fig.~\ref{fig:Fe_lines}), but we note that the lower effective area of the MEG detector towards those energies resulted in lower signal-to-noise detection of these lines. On the other hand, the {\it Chandra} HEG spectrum unambiguously resolves the three Fe lines. A similar modeling approach to that performed for the MOS spectrum was done for the MEG and HEG data and the results are also listed in Table~\ref{tab:Fe_lines}.

\subsection{The high-resolution spectra}
\label{sec:high-res-rgs}

The high-resolution X-ray spectra of CH\,Cyg presented in Fig.~\ref{fig:rgs_spec} show the contribution from a number of emission lines in the 1--36~\AA\, wavelength range, which corresponds to the $\sim$0.3--7~keV energy range. Their origin can be attributed to emission lines from the C\,{\sc vi}, N\,{\sc vi}, N\,{\sc vii}, O\,{\sc vii}, O\,{\sc viii}, Ne\,{\sc ix}, Ne\,{\sc x}, Al\,{\sc xii}, Mg\,{\sc xi}, Mg\,{\sc xii}, Si\,{\sc xiii}, Si\,{\sc xv}, Si\,{\sc xvi}, Fe\,{\sc xvii} and Fe\,{\sc xviii} ions. It is important to note that the RGS spectrum is mainly sensitive to soft X-ray emission ($\lambda>$10~\AA) and the MEG and HEG spectra are more sensitive to higher energies ($\lambda<$10~\AA).

A list of emission lines present in the RGS, MEG and HEG spectra is presented in Table~\ref{tab:rgs_lines}. These were measured by modelling each spectral feature with Gaussian fits in {\sc xspec}. In order to estimate the line fluxes, the local continuum around the emission lines was modelled adopting power law components. This is specially important for the MEG and HEG spectra as illustrated in Fig.~\ref{fig:rgs_spec}. Table~\ref{tab:rgs_lines} shows the expected ($\lambda_\mathrm{lab}$) and measured wavelengths ($\lambda_\mathrm{obs}$), the energy $E$, the line width $\sigma$, $EW$ and the total observed flux $f$ per line for lines with signal-to-noise levels larger than 3.

The high-resolution spectra can be used to assess the multi-temperature nature of the X-ray emission material and as the best tool for determining the abundances. First, we can assess the temperature of the X-ray-emitting plasma from the high-resolution spectra by using the He-like triplets. The He-like ions are composed by a resonance ($r$), an intercombination ($i$) and a forbidden ($f$) line. The ratio of their different components can be used to estimate the physical properties of hot gas \citep[see][and references therein]{Porquet2000}. The $G$ ratio, which is a function of the electron temperature ($T_\mathrm{e}$), is defined as
\begin{equation}
    G(T_\mathrm{e}) = \frac{f + i }{r}.
\end{equation}
\noindent This can be calculated using the fluxes of each emission line in a triplet.

The wavelength range of the high-resolution spectra analysed here should include the N\,{\sc vi} 28.78,29.08,29.53~\AA, O\,{\sc vii} 21.60,21.80,22.10~\AA, Ne\,{\sc ix} 13.45,13.55,13.70~\AA, Mg\,{\sc xi} 9.17,9.23,9.31~\AA\, and Si\,{\sc xiii} 6.65,6.69,6.74~\AA\, He-like triplets. In the present observations we detect all of those of N\,{\sc vi}, O\,{\sc vii}, Mg\,{\sc xi} and Si\,{\sc xiii} 
with sufficient signal to noise (see Fig.~\ref{fig:rgs_spec}) to resolve each of the components and, consequently, to draw temperature estimates.

\begin{figure*}
\begin{center}
  \includegraphics[angle=0,width=0.45\linewidth]{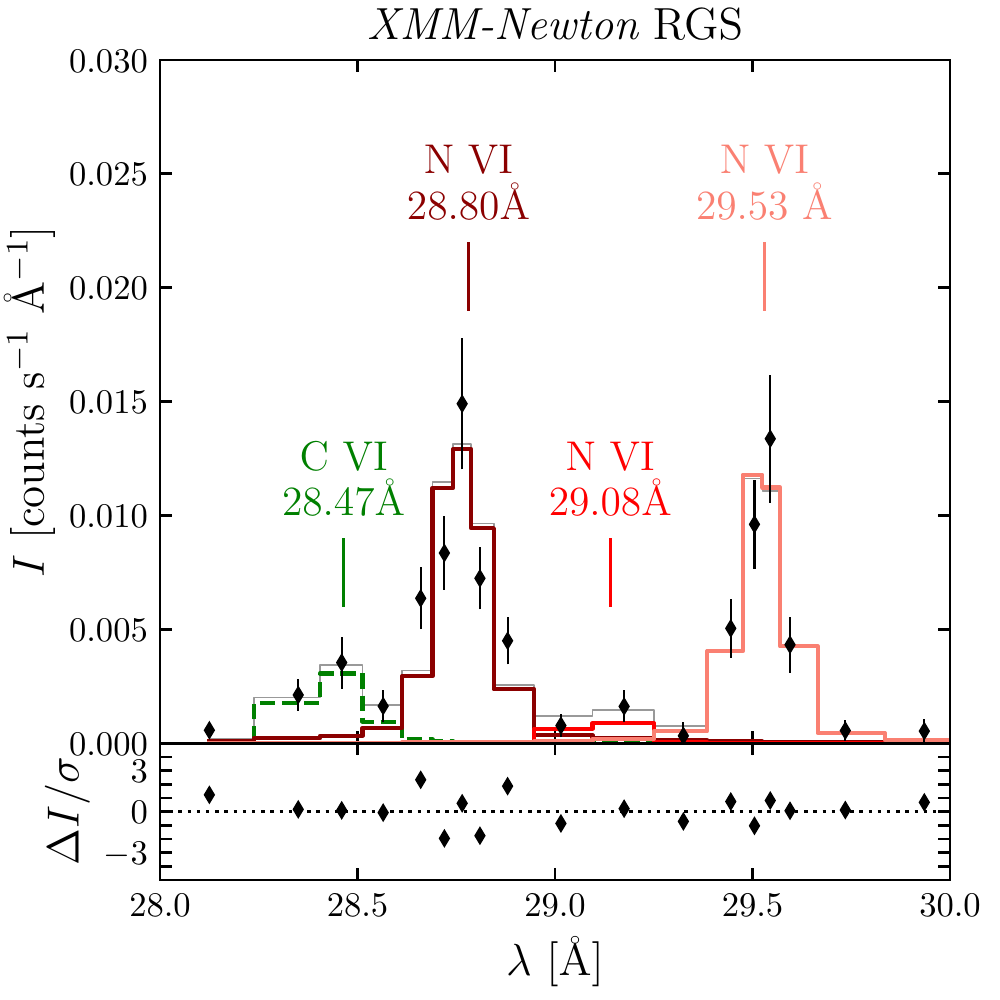}~  \includegraphics[angle=0,width=0.45\linewidth]{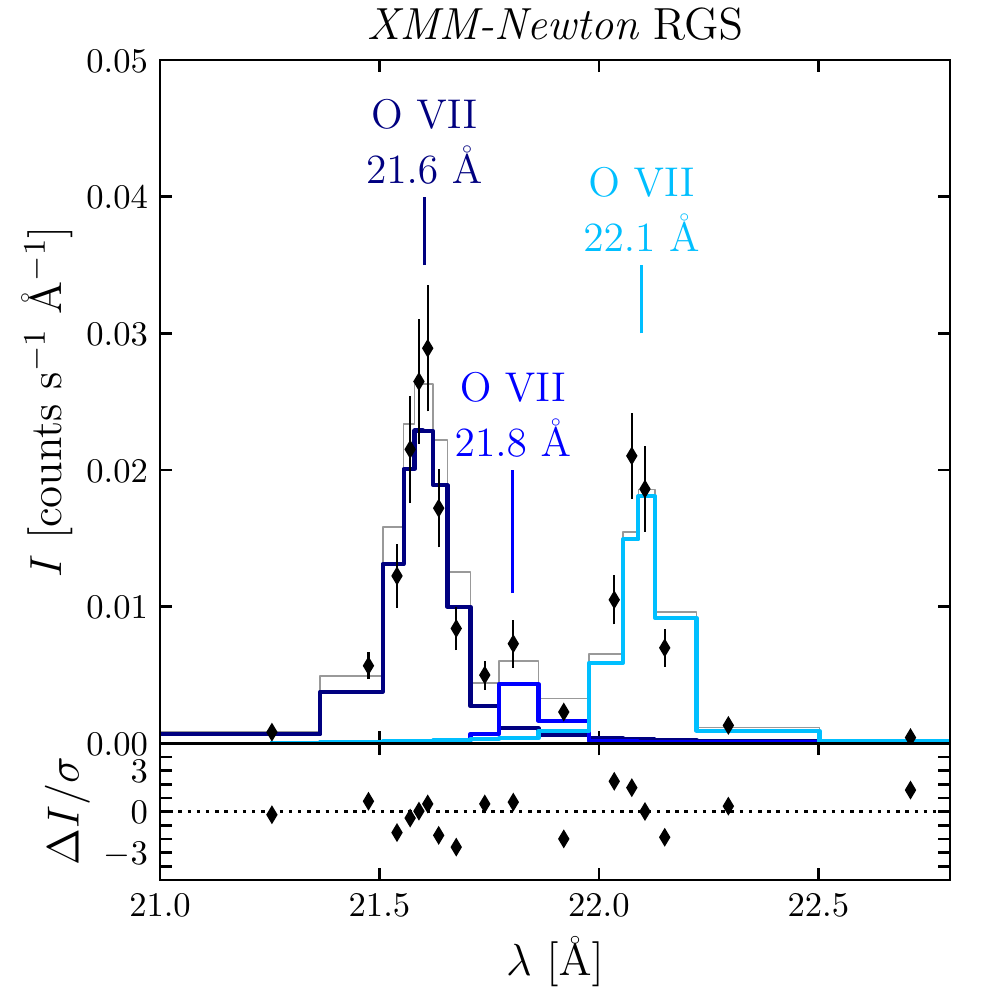}\\
  \includegraphics[angle=0,width=0.45\linewidth]{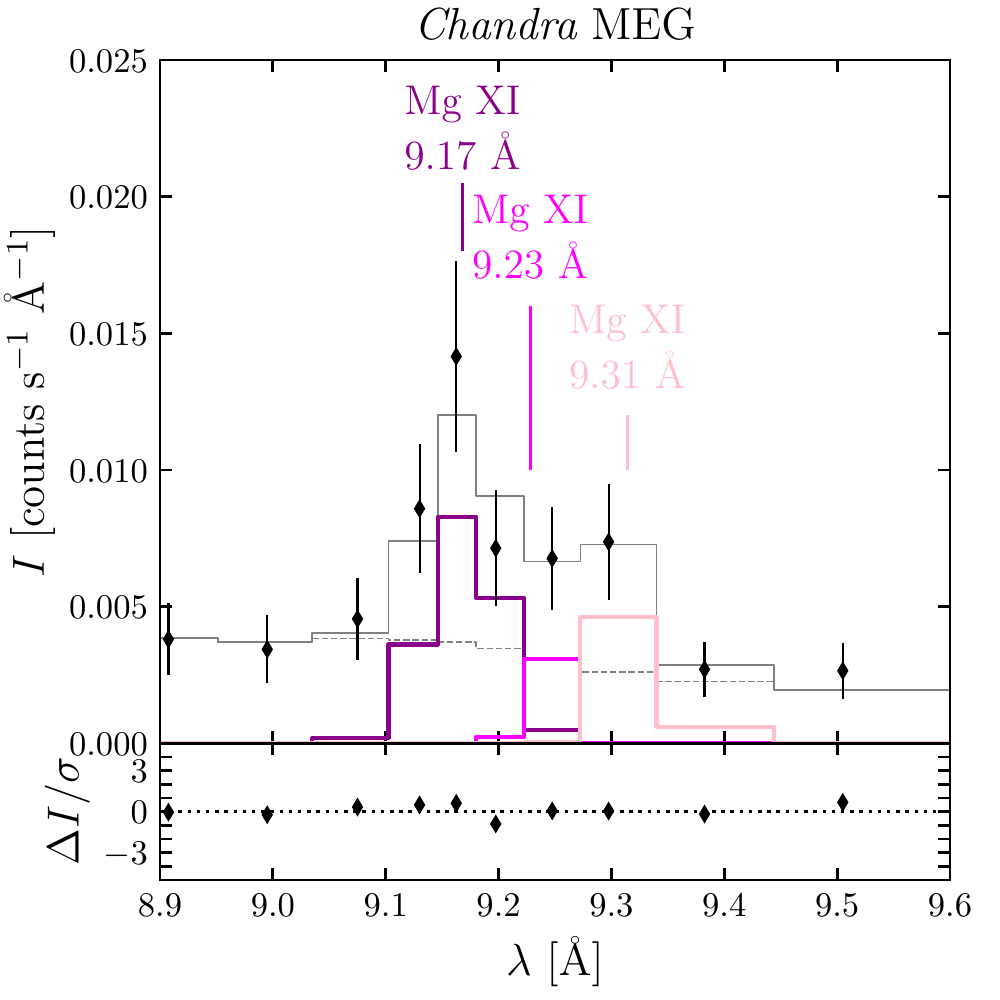}~
  \includegraphics[angle=0,width=0.45\linewidth]{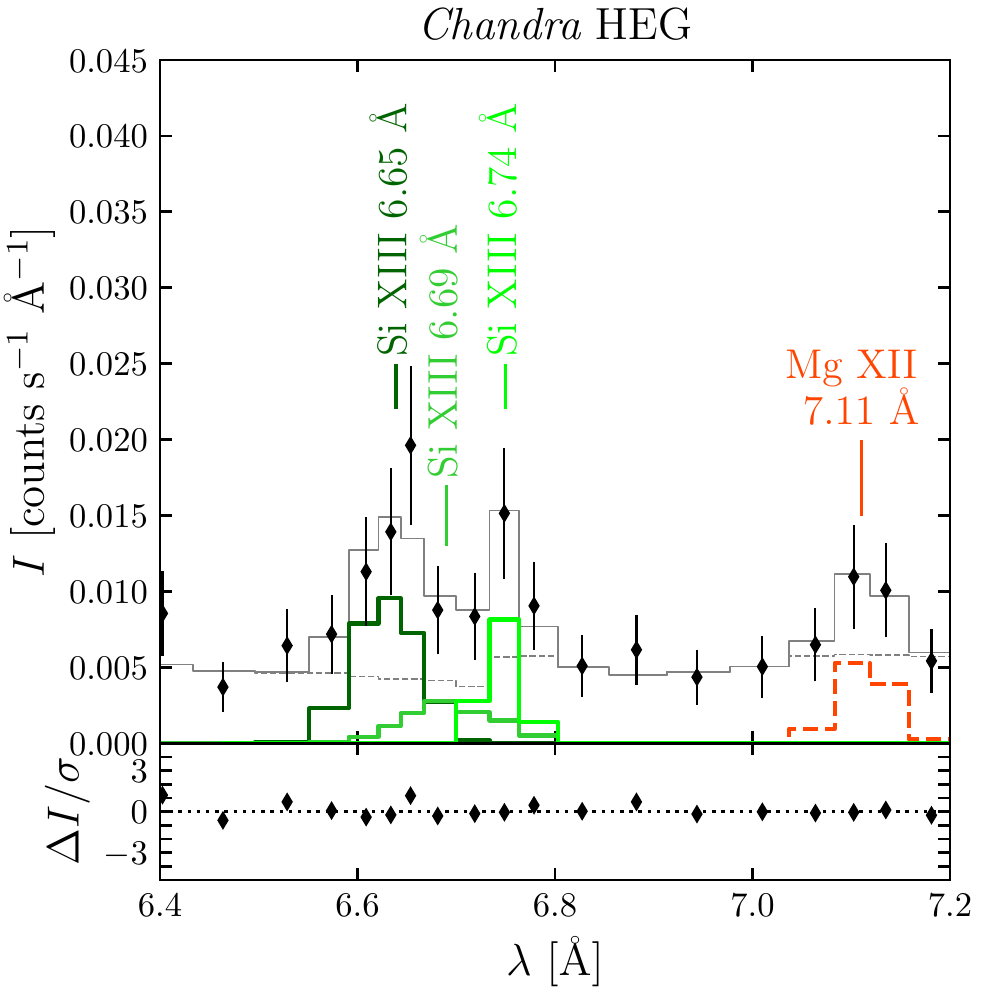}
\end{center}
\caption{Details of the N\,{\sc vi} (top left), O\,{\sc vii} (top right), Mg\,{\sc xi} (bottom left) and Si\,{\sc xiii} (bottom right) He-like triples detected in the high-resolution X-ray observations analysed in the present paper. Fits to the individual emission lines are shown with different colours and wavelengths. In the case of the N\,{\sc vi} triplet we also show the contribution of the C\,{\sc vi} line at 28.47~\AA. For the Mg\,{\sc xi} and Si\,{\sc xiii} triplets a local continuum was modelled and subtracted. Residuals between the total models and the data are also shown.}
\label{fig:O_lines}
\end{figure*}

Fig.~\ref{fig:O_lines} presents the details of the N\,{\sc vi}, O\,{\sc vii}, Mg\,{\sc xi} and Si\,{\sc xiii} He-like triplets showing the best fits to their emission lines. Particular care was taken to subtract the contribution of the C\,{\sc vi} 28.47~\AA\, line towards the lower wavelength range of the N\,{\sc vi} triplet (see Fig.~\ref{fig:O_lines} top left panel). For the case of the Mg\,{\sc xi} and Si\,{\sc xiii} triplets the local continuum has a significant contribution and had to be modelled in order to estimate appropriate line fluxes.

We used the flux estimates listed in Table~\ref{tab:rgs_lines} and calculated $G$ values of 0.62$\pm$0.18, 0.83$\pm$0.19, 1.07 and 1.0 for the N\,{\sc vi}, O\,{\sc vii}, Mg\,{\sc xi} and Si\,{\sc xiii} He-like triplets, respectively. We note that the errors of the $G$ parameter for the Mg\,{\sc xi} and Si\,{\sc xiii} triplets are exceedingly large \citep[outside the validity range; see][]{Porquet2000}, thus hampering a reliable temperature range for these two triplets. We will present their values, but we note that strong conclusions from the Mg\,{\sc xi} and Si\,{\sc xiii} triplets is not advised.

Using the approach presented in \citet{Porquet2000} and assuming that the physical origin of the soft X-ray emission from CH\,Cyg is due to shocks, we estimate electron temperatures of 
$T_\mathrm{e}$(N\,{\sc vi}) = $(2.3\pm0.7)\times10^{6}$,
$T_\mathrm{e}$(O\,{\sc vii}) = $(2.4\pm0.6)\times10^{6}$,
$T_\mathrm{e}$(Mg\,{\sc xi}) = 4.6$\times10^{6}$
and
$T_\mathrm{e}$(Si\,{\sc xiii})=7.5$\times10^{6}$~K.

To obtain reliable abundance estimations for the X-ray-emitting material, we need to model the high-resolution spectra. To obtain the best fit to the RGS spectrum we followed a somewhat iterative process similarly to that described for the MOS data. We started by using a single plasma temperature model, but this was not sufficient to obtain an acceptable fit to the RGS spectrum. A two-temperature plasma model improved the fit but had some difficulty reproducing the N lines. In addition, the lower wavelength range of the spectrum ($5 < \lambda < 15$~\AA) was underestimated. A final three-component model best reproduced the emission lines in the RGS spectrum. We found that the fit further improved by letting the C, N, O and Fe abundances vary.

An acceptable fit, which is labelled as Model D, resulted in similar components as those for the soft energy range of the MOS spectrum (see Table~\ref{tab:RGS_par}): two plasma components of $kT_1=$0.12 and $kT_2=$0.47 with the contribution of a power law. The best-fit absorbing column density is $N_\mathrm{H}=(4\pm0.2)\times10^{20}$~cm$^{-2}$. Model D is compared with the observed RGS spectrum in the top panel of Fig.~\ref{fig:rgs_spec}. That panel shows that Model D makes a good job reproducing the intensity of most emission lines in the RGS spectrum. Similarly to the spectral analysis of the MOS data, the third component can also be exchanged by a third optically-thin emission plasma model with similar statistics. The later is labelled as Model E and is compared to Model D in Table~\ref{tab:RGS_par}.

Both the MEG and HEG spectra were modelled with absorbed two-component plasma models with a Gaussian representing the Fe fluorescent line. The column density and temperatures are higher than those obtained for the RGS spectra but only because the MEG and HEG data are likely to trace the heavily-extinguished X-ray-emitting material that includes the accretion disk. Solar abundances were sufficient for producing acceptable spectral fits to both the MEG and HEG spectra. The details of models F and G, which correspond to the best-fit data of the MEG and HEG spectra, respectively, are presented in Table~\ref{tab:RGS_par} and are compared with the observed spectra in the middle and bottom panels of Fig.~\ref{fig:rgs_spec}.

\begin{table}
\begin{center}
\setlength{\columnwidth}{0.1\columnwidth}
\setlength{\tabcolsep}{0.7\tabcolsep}
\caption{Details of out best models to the high-resolution X-ray spectra of CH Cyg.}
\label{tab:RGS_par}
\begin{tabular}{lcccc}
\hline
parameter                             & Model D           & Model E            & Model F            & Model G      \\ 
                                      & RGS               & RGS                & MEG                & HEG          \\  
                                      & (5--37~\AA)       & (5--37~\AA)        & (1.5--15~\AA)      & (1.5--15~\AA)\\  
\hline
$\chi^2$   & 447.00/181                                   & 447.62/181         & 419/410            & 352.01/372   \\
$N_\mathrm{H,1}$ [10$^{22}$~cm$^{-2}$]& 0.04$\pm$0.02     & 0.04$\pm$0.02      & 1.2$\pm$0.1        & 1.0$\pm$0.2  \\
$kT_1$          [keV]                 & 0.12$\pm$0.01     & 0.12$\pm$0.01      & 1.05$\pm$0.03      & 1.11$\pm$0.10 \\
$A_1$           [cm$^{-5}$]           & 1.8$\times10^{-3}$& 1.7$\times10^{-3}$ & 2.4$\times10^{-3}$ & 2.4$\times10^{-3}$\\
$N_\mathrm{H,1}$ [10$^{22}$~cm$^{-2}$]& \dots             & \dots              & 5.3$\pm$0.2        & \\
$kT_2$          [keV]                 & 0.47$\pm$0.03     & 0.46$\pm$0.02      & 8.0                & 8.42$\pm$0.30 \\
$A_2$           [cm$^{-5}$]           & 1.0$\times10^{-4}$& 5.7$\times10^{-4}$ & 3.7$\times10^{-2}$ & 3.7$\times10^{-2}$\\
$\Gamma$                              & 0.84              & \dots              & \dots              & \dots \\
$kT_3$          [keV]                 & \dots             & 42$\pm$200         & \dots              & \dots \\
$A_3$           [cm$^{-5}$]           & 1.1$\times10^{-4}$& 7.4$\times10^{-4}$ & \dots              & \dots \\
$E_\mathrm{line}$[keV]                & \dots             & \dots              & 6.4                & 6.40  \\
$\sigma$ [keV]                        & \dots             & \dots              & 4.2$\times10^{-2}$ & 4.2$\times10^{-2}$\\
$A_\mathrm{line}$ [cm$^{-5}$]         & \dots             & \dots              & 1.8$\times10^{-4}$ & 1.1$\times10^{-4}$\\
\hline
C    & 1.2$\pm$0.3   & 1.2$\pm$0.3   & 1.0 & 1.0 \\
N    & 2.8$\pm$0.4   & 2.8$\pm$0.5   & 1.0 & 1.0 \\
O    & 1.4$\pm$0.2   & 1.4$\pm$0.2   & 1.0 & 1.0 \\
Mg   & 1.0           & 1.0           & 1.0 & 1.0 \\
Si   & 1.0           & 1.0           & 1.0 & 1.0 \\
Fe   & 0.3$\pm$0.1   & 0.3$\pm$0.1   & 1.0 & 1.0 \\
\hline
$f$ [erg~s$^{-1}$~cm$^{-2}$] &  1.9$\times10^{-12}$ & 2.0$\times10^{-12}$ & 3.3$\times10^{-11}$ & 3.3$\times10^{-11}$\\
$F$ [erg~s$^{-1}$~cm$^{-2}$] &  2.5$\times10^{-12}$ & 2.6$\times10^{-12}$ & 6.5$\times10^{-11}$ & 6.8$\times10^{-11}$\\
\hline
\end{tabular}
\end{center}
\end{table}

\section{Discussion}
\label{sec:diss}

\subsection{General properties of the X-ray emission of CH\,Cyg}
\label{sec:model_disk}

Previous studies of medium-resolution X-ray spectra of CH\,Cyg suggested the presence of slightly-absorbed two-temperature plasma components plus a hotter and heavily-absorbed thermal component likely due to the accretion disk (see references in Sec.~\ref{sec:intro}). We have found similar components in our study but also require  an additional component to reproduce the emission in the 2.0--4.0~keV energy range. Three different models with an {\it apec}, power law and a reflection component were attempted in Sec.~\ref{sec:epic-spec}. Although these models produce statistically-acceptable models, they might not be useful to explain the physics behind the production of X-rays in that wavelength range.

We suggest that the optimal physical explanation is the presence of a reflection component as it naturally explains the excess of emission in the 2.0--4.0 keV energy range and is expected to be correlated with the Fe K$\rm{\alpha}$ emission line (see Model C in Fig.\,\ref{fig:spec} and Section\,\ref{sec:epic-spec}). 
Indeed, the reflection component is common in AGNs and is mainly composed of two features: i) the Compton hump with an emission peak associated to the column density of the reflector and ii) the Fe K$\alpha$ emission line attached to the Compton-hump \citep[see, e.g.][and references therein]{Osorio-Clavijo22}. In AGNs the reflection component is mainly attributed to an ionised reflector associated to the accretion disk (with relativistic effects) and/or a distant and neutral reflection associated to the torus \citep[see][for a review]{Ramos-Almeida17}. The shape of this reflection component depends on the chemistry, ionisation state, geometry and viewing angle toward the system \citep{Fabian05}. In the AGN field several models for the reflection have been proposed \citep[see for example][]{Balokovic2018,Garcia14}. 

\begin{figure}
\begin{center}
  \includegraphics[angle=0,width=\linewidth]{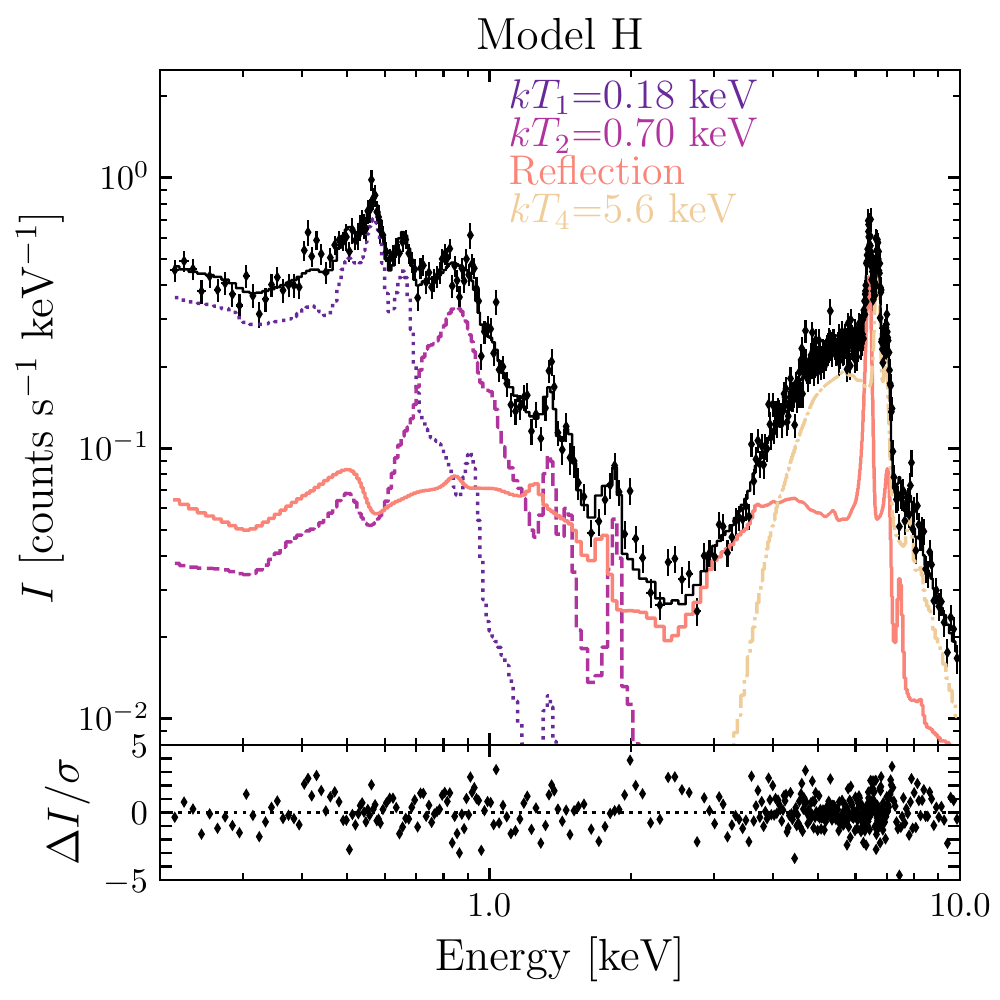}
\end{center}
\caption{Left: {\it XMM-Newton} MOS (MOS1+MOS2)
  spectrum of CH\,Cyg compared to the best model obtained by adopting an ionised disk as a reflection component (Model H) described in Sec~\ref{sec:model_disk}. The black diamonds correspond to the observed spectrum while the black histogram represents the best-fit model with the bottom panels showing the residuals. Different model components are shown with different lines (colours). Right: A close up of the Fe emission lines.}
\label{fig:spec_final}
\end{figure}

To further peer into the properties of the hard X-ray emission from CH\,Cyg we produced similar models as those for AGNs. In Section~\ref{sec:epic-spec} we showed that the {\it borus09} model described by \citet{Balokovic2018}, which assumes a smooth distribution of neutral gas, fails to reproduce the MOS data. This is probably due to the fact that the reflector in CH\,Cyg is likely an ionised disk. In order to test this hyphothesis, we ran new radiative transfer simulations using the code {\sc RefleX}\footnote{{\sc RefleX} executable, user manual and example models are available at \url{http://www.astro.unige.ch/reflex}} \citep{Paltani2017}. {\sc RefleX} simulates the physical processes of propagation of X-rays through matter around the central source using Monte Carlo simulations to track individual photons. It includes several spectral shapes for the incident radiation and geometries for the reflector that can be built in blocks to create complex geometries. We adopted a power-law energy distribution (with a photon index $\rm{\Gamma=2}$) for the incident radiation from a point-like source. The luminosity in the 0.2--10.0~keV energy range of the incident radiation was fixed to $L_\mathrm{X}=2\times 10^{33}$ erg~s$^{-1}$ (see Table\,\ref{tab:EPIC_par}).

We adopt a disk geometry as the likely cause of reflection in CH\,Cyg. The model parameters include the disk outer radius ($r_\mathrm{out}$) and thickness ($h_\mathrm{thick}$). We ran several models and found that the resultant spectrum is not very sensitive to $r_\mathrm{out}$ since most of the reflection occurs at the inner region. We explored several disk thickness as a fixed fraction of the outer radius in the  $h_\mathrm{thick}$=[0.001--0.2]~$r_\mathrm{out}$ range. We also explored several column densities of the disk in the [4--7]$\times 10^{23}$~cm$^{2}$ range, which is consistent with our spectral analysis (see Table\,\ref{tab:EPIC_par}). Finally, we also fixed the abundances to those determined from our spectral analysis of the RGS spectrum (see Table\,\ref{tab:RGS_par}).

A total of 40 base models were performed. For each of them we created 10 spectral energy distributions (SEDs) with viewing angles between 0 and 90$\rm{^\circ}$. We converted these models to an additive one-parameter table (in fits format) associated with all the SEDs using the {\it flx2tab} task within {\sc heasoft}\footnote{\url{https://heasarc.gsfc.nasa.gov/docs/software/lheasoft/}}. We then used {\sc xspec} to test our new reflection models by replacing the {\it pexrav} and Fe K$\alpha$ emission line Gaussian components. The result is shown in Fig.~\ref{fig:spec_final} and is labelled as Model H. The reflection component dominates the 2.0--4.0 keV energy range and it naturally reproduces the strength of the Fe K$\alpha$ emission line without the need for the Gaussian components. Furthermore, although not dominant, the reflection component also contributes to energies below 2.0 keV. The reduced $\rm{\chi^2 = 1.28}$, which is slightly better than the models to the MOS data presented in Sec.~\ref{sec:epic-spec}. The optimal parameters for the reflector component based on our analysis are $N_\mathrm{H}=5\times 10^{23}$~cm$^{-2}$ and $h_\mathrm{thick}=0.1 r_\mathrm{out}$. However, a full grid of models is required to properly explore the parameters of the model. Future results and fits to several $\beta/\delta$-type sources will be presented in a separate work.

The temperatures and abundances of the two thermal components associated with the soft X-ray emission are fully consistent with other models of the MOS data (see Table~\ref{tab:EPIC_par} and Fig.~\ref{fig:spec_final}). The line-of-sight column density and the temperature associated to the thermal component contributing to the hard X-rays show slightly different values of $N_\mathrm{H}=(7.5\pm 0.2)\times 10^{23}$~cm$^{-2}$ and $kT = 5.6\pm0.1$~keV. The results of our analysis suggests that reflection from the ionised disk is relevant in SySts. This is similar to the presence of reflection components in non-AGN sources, particularly Galactic black-hole binaries GX\,339-4 by \citet{Miller04} and XTE\,J1650-500 by \citet{Miniutti2004} \citep[see also][for a review]{Reynolds13}.

\begin{figure}
\begin{center}
  \includegraphics[angle=0,width=\linewidth]{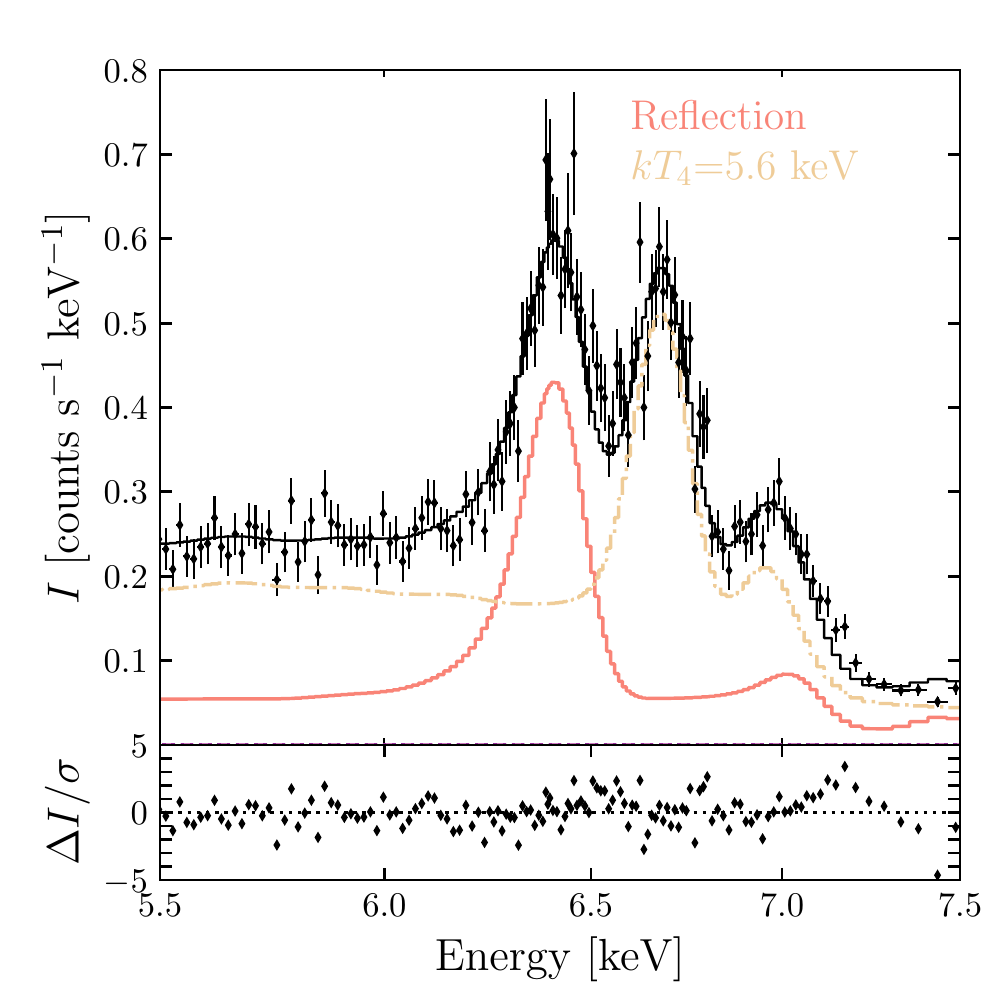}
\end{center}
\caption{Same as Fig.~\ref{fig:spec_final} but for the 5.5--7.5~keV energy range that shows the 6.4, 6.7 and 6.97~keV Fe emission lines.}
\label{fig:spec_final_Fe}
\end{figure}

The reflection component in CH\,Cyg naturally produces the 6.4~keV Fe line and a non-negligible contribution to the 6.97 keV emission line due to the Compton shoulder of the line. The heavily-obscured plasma component (in our case that of 5.6~keV) produces the 6.7~keV Fe line and most of the emission at 6.97~keV. This situation is illustrated in Fig.~\ref{fig:spec_final_Fe}. Gaussian fits to these emission lines are typically performed when studying SySt and other accreting systems with hard X-ray emission \citep[such as AGNs and cataclysm variables; see, e.g.,][]{Esaenwi2015,Eze2015,Danehkar2021,Dutta2022,GonzalezMartin2011}, but strictly speaking, single Gaussian components are not based in a physical process. The reflection component accurately predicts the emission and provides new insight on the physical processes in the system.  We suggest that further tailoring of reflection component models to $\beta/\delta$-type SySts is a promising avenue of future exploration.

\citet{Wheatley2006} were the first to suggest that the soft X-ray emission from CH\,Cyg could be interpreted as scattering of the hard X-ray source from an ionised medium. Although their model seems to reproduce a similar strength of the soft X-ray emission from the observations, is does not reproduce the spectral shape of the soft emission for $E<$2.0~keV.
The reflection model presented here is commonly used in AGN X-ray analysis. Our spectral model suggests that the reflection component contributes but does not dominates the soft X-ray regime. On the other hand, the soft X-ray emission is best reproduced with optically-thin emission plasma models, which explain the evident emission lines in the soft energy range ($E<$2~keV). We attribute the presence of the soft emission to the extended component (very likely jets) which has been detected using {\it Chandra} \citep[][]{Karovska2007,Karovska2010}. A similar scenario can be envisaged in R~Aqr where extended X-ray emission has been mapped in detail by {\it XMM-Newton} and {\it Chandra} \citep[see][and references therein]{Toala2022}.

Preliminary analysis of PSF deconvoluted images of the 2008 {\it Chandra} ACIS-S3 data presented in \citet{Karovska2010} show that the reflector component (traced by the 2.0--4.0~keV) is mainly distributed within the inner 0.5~arcsec. This suggests that this emission is likely originating from the binary (WD-accretion disk) itself. For comparison we note that \citet{Mukai2007} estimated that the central engine should be contained within the inner 0.4~arcsec ($\approx$80 AU at a distance of 200 pc). Further analysis of such structures will be pursued in a subsequent paper.

\subsection{On the variability of the X-ray emission}

CH\,Cyg has been known to be a variable X-ray source for decades \citep[see, e.g.,][]{Luna2020}. The analysis of multi-epoch, multi-mission X-ray observations of CH\,Cyg presented by \citet{Mukai2007} showed that between 1994 and 2006 its hard X-ray flux was reduced more than an order of magnitude. The observed flux of the hard band was reported to be about 60$\times10^{-12}$~erg~s$^{-1}$~cm$^{-2}$ in the 1994 {\it ASCA} data and 2$\times10^{-12}$~erg~s$^{-1}$~cm$^{-2}$ in the 2006 {\it Suzaku} era. The analysis of the {\it Chandra} and {\it XMM-Newton} data presented here shows that the flux has since increased to $\gtrsim$30$\times10^{-12}$~erg~s$^{-1}$~cm$^{-2}$ in 2009 (see Table~\ref{tab:RGS_par}) and to even larger fluxes in 2018 ($>$70$\times10^{-12}$~erg~s$^{-1}$~cm$^{-2}$; see Table~\ref{tab:EPIC_par}). The flux in 2018 is similar to the value reported for the {\it ASCA} measurements obtained 24~years ago.

Variability has also been reported from the Fe lines. Based on the available X-ray data of CH\,Cyg, this variability does not seem to be related to the flux of the heavily absorbed source. Previous results on the analysis of the Fe lines (see references in Sec.~\ref{sec:intro}) showed that the dominance of the Fe fluorescent line at 6.4~keV has changed dramatically. From a low contribution in the 1994 {\it ASCA} spectrum, a negligible contribution in the 2001 {\it Chandra} data, to a complete dominance in the 2006 {\it Suzaku} observation. In the 2009 {\it Chandra} HEG (this work), the fluorescent 6.4~keV line has about a third of the flux of the 6.7~keV emission line, while in the 2018 {\it XMM-Newton} MOS spectrum suggests these two lines have similar fluxes.

The fluorescent Fe line at 6.4~keV is produced by irradiation of surrounding material of a hard X-ray source. Therefore the $EW$ can be related with the covering angle of the absorbing material $\Omega$ producing the fluorescent Fe emission line and the column density $N_\mathrm{H}$ as \citep[see][]{Inoue1985}
\begin{equation}
    EW = \left(\frac{\Omega}{4 \pi}\right) \left( \frac{N_\mathrm{H}}{10^{21}~\mathrm{cm}^{-2}} \right)~\mathrm{eV}.
\end{equation}
In particular, for the case of a SySt the absorbing matter might be a dense accretion disk, the dense wind from the red giant companion or a combination of both. \citet{Eze2014} suggested that for CH\,Cyg the fluorescent line is mostly produced in the accretion disk as it is irradiated by hard X-ray emission produced at the boundary layer between the accretion disc and the WD. Model H seems to corroborate this idea, demonstrating that the 6.4~keV Fe line can be produced as a consequence of reflection from the ionised disk. 
It seems that the dramatic changes in $EW$ of the fluorescent line suggest that the accretion disk is not a uniform steady structure but a highly variable. The reduction of the $EW$ of the fluorescent line in the {\it Chandra} HEG observations is consistent with the reduction in column density needed to fit this spectrum.

\subsection{High-resolution spectra and abundances}

The RGS, MEG and HEG spectra of CH~Cyg presented here are the first high-resolution grating X-ray spectroscopy of this SySt. Individual analysis of the RGS, MEG and HEG spectra seem to reproduce the properties of the X-ray-emitting plasma detected with {\it XMM-Newton} MOS and previous missions in a complementary way. The best models of the RGS data are consistent with the components required to fit the MOS data in the soft X-ray range. The latter is dominated by plasma with temperatures of 0.12~keV(=1.4$\times10^{6}$~K). The MEG and HEG spectra are most sensitive to the heavily absorbed part of the X-ray spectrum of CH\,Cyg, which corresponds to the higher-temperature plasma associated with the accretion disk.

\begin{table}
\setlength{\columnwidth}{0.1\columnwidth}
\setlength{\tabcolsep}{0.8\tabcolsep}
\caption{Photospheric abundances reported for the red giant component in CH\,Cyg taken from \citet[][]{Schmidt2006} compared with the abundance estimates from the Models A and D. Solar abundances are adopted from \citet{Lodders2009}.}
\begin{center}
\begin{tabular}{ccccc}
\hline
        & \multicolumn{2}{c}{\citet{Schmidt2006}} & RGS (Model D)     & MOS (Model A)  \\ 
Element & 12+log$X$   & $X/X_{\odot}$             & $X/X_{\odot}$ & $X/X_{\odot}$  \\  
\hline
\vspace{1mm}
C  & 8.37$\pm$0.22  & 0.95$^{+0.63}_{-0.38}$      & 1.2$\pm$0.3       & 1.0 \\
\vspace{1mm}
N  & 8.08$\pm$0.13  & 1.65$^{+0.58}_{-0.43}$      & 2.8$\pm$0.4       & 2.9$\pm$0.5    \\
\vspace{1mm}
O  & 8.76$\pm$0.24  & 1.07$^{+0.79}_{-0.45}$      & 1.4$\pm$0.2       & 0.8$\pm$0.1    \\
\vspace{1mm}
Mg & 8.68$\pm$0.21  & 13.08$^{+8.58}_{-5.29}$     & 1.0               & 6.3$\pm$2.2    \\
\vspace{1mm}
Si & 7.40$\pm$0.18  & 0.74$^{+0.38}_{-0.25}$      & 1.0               & 0.5$\pm$0.5    \\
\vspace{1mm}
Fe & 7.50$\pm$0.19  & 1.09$^{+0.60}_{-0.38}$      & 0.3$\pm$0.1       & 0.9$\pm$0.2    \\ 
\hline
\end{tabular}
\end{center}
\label{tab:abund}
\end{table}

The He-like triplets resolved by the high-resolution spectra were used to estimate the plasma temperatures. Under the assumption that the X-ray emission is produced solely by shocks, the resolved N\,{\sc vi}, O\,{\sc vii}, Mg\,{\sc xi} and Si\,{\sc xiii} triplets suggest plasma temperature of 2.3$\times10^{6}$~K (=0.20~keV), 2.4$\times10^{6}$~K (=0.21~keV), 4.6$\times10^{6}$~K (=0.40~keV) and 7.5$\times10^{6}$~K (=0.65~keV), respectively. Within their error bars, these line-based plasma temperatures agree with those required to fit the entire X-ray spectra.

The spectral fitting of the MOS and RGS spectra presented here allowed us to perform abundance determination of the X-ray-emitting gas. In particular, we note that the abundances required to reproduce the RGS spectrum (that of the soft X-ray emission) are very close to those determined for the cool stellar component in CH\,Cyg. The best-fit to the RGS spectrum determines C, N and O abundance values that are similar to those reported in \citet{Schmidt2006} for CH\,Cyg. Table~\ref{tab:abund} compares the estimated values from optical observations with those obtained from the RGS spectral fit (Model D). We note that analysis of the RGS spectrum was unable to constrain Mg and Si abundances, while the Fe abundance is constrained to about 0.3 times the solar value, whilst that estimated from \citet{Schmidt2006} is about solar. For further comparison, we also list in Table~\ref{tab:abund} the abundances estimates from Model A performed to the MOS spectrum. Within error bars, the values are consistent with estimates from \citet{Schmidt2006}. This demonstrates that the X-ray-emitting material has a chemical composition similar to that of the cool companion and is likely shock heated when the material is accreted and ejected by the WD.

On the other hand, good quality fits to the MEG and HEG spectra (the harder part of the X-ray emission from CH\,Cyg) were achieved by adopting solar abundances. These fits resulted in reduced $\chi^{2}$ of $\sim1$ and we preferred not to further explore other abundance values. However, we note that the MEG and HEG are not very informative to the abundances of elements like C, N and O.

\subsection{Comparison with other sources}
\label{sec:other_sources}

We have shown through our spectral analysis that CH\,Cyg shares very similar emission characteristics with AGNs, an interesting approach initially discussed by \citet[][]{Wheatley2006} based on the {\it ASCA} X-ray spectrum of this SySt and comparing it to those of Seyfert 2 galaxies. The similarities are undeniable and evidently reflect the common physical processes behind the production of X-ray radiation around accreting compact sources. It is likely that some SySts are more active (higher accretion rates) than others and this might explain the diverse zoo of X-ray-emitting  properties, as is the case for AGN \citep[see][and references therein]{Merc2019}.

Most SySt detected in X-rays have lower-quality spectra, nevertheless, some of them have X-ray spectra that resemble those reported for CH\,Cyg: a considerable contribution to soft energies ($E<$2~keV) with a clear detection of emission lines plus a heavily-extinguished hard component ($E>2$~keV) with the contribution from the Fe lines \citep[e.g.,][and references therein]{Luna2013}. For example, NQ\,Gem, UV\,Aur and ZZ\,CMi discussed in \citet{Luna2013} and \citet{Toala2023}. There is no doubt that the hard emission from those $\beta/\delta$-type systems is also produced by their accretion disks and reflection physics as demonstrated here for CH~Cyg. Needless to say, SySts are a scaled-down version of accreting super massive black holes.

\citet{Karovska2010} resolved the different components in CH\,Cyg using {\it Chandra}, {\it HST} and VLA observations. These multiwavelength observations were used to spatially-resolve the jets and central source, and allowed independent spectral analysis of the different components.
The brightest X-ray-emitting clump in the jet exhibits spectral characteristics that are significantly different than that of the central source but at comparable flux values \citep[see fig.~4 in][]{Karovska2010}. This confirms that a fraction of the soft X-ray emission detected in the MOS data is due to shocks produced by the jet activity.

A similar situation can be observed in the well-resolved SySt, R~Aqr. It has been spatially-resolved and, consequently, one can directly associate  spectral characteristics to different morphological features. {\it Chandra} observations of R~Aqr have shown that the central stellar system is associated with the hard X-ray emission produced by the accretion disk \citep{Kellogg2001}, but soft X-ray emission peaking at $\lesssim$0.6~keV is associated with the jet-like structures protruding from R~Aqr \citep{Kellogg2007}. Recently, \citet{Toala2022} unveiled the presence of extended ($\sim$2~arcmin) soft emission detected with {\it XMM-Newton} observations with dominant plasma temperature of $\sim$0.1~keV.

We used the available {\it XMM-Newton} EPIC observations of R~Aqr to extract a spectrum that encompasses all structures in this SySt. The resultant spectrum and best fit model are presented in Appendix~\ref{sec:app} and the last column of Table~\ref{tab:EPIC_par}. The spectral similarities between R~Aqr and CH\,Cyg suggest that SySt classified as part of the $\beta/\delta$ sequence might be a combination of spatial distinct features that are often unresolved. This hypothesis will be further investigated with future simulations.

\section{Conclusions}
\label{sec:conclusions}

We reported the analysis of publicly available {\it XMM-Newton} and {\it Chandra} observations of the SySt CH\,Cyg. Our analysis of these observations includes the first study of {\it XMM-Newton} RGS and {\it Chandra} HETG high-resolution spectra ever presented for this SySt. The analysis of the medium-resolution MOS X-ray spectra of CH\,Cyg are consistent with the components suggested by the modeling of the high-resolution spectra and by previous studies, however, the optimal model requires an additional component to fit the 2.0--4.0~keV energy range. The temporal variation of the X-ray properties were also discussed. Our results can be summarised as follows:

\begin{itemize}

\item The analysis of the medium-resolution spectra derived from the MOS observations require an additional component to improve the fit in the 2.0--4.0~keV energy range. Several components were attempted, but we conclude that the optimal component is that of a reflection component. We borrowed  techniques used to model reflection in AGN to determine that reflection is produced by the inner parts of an ionised accretion disk in CH~Cyg. This reflection component is able to simultaneously explain the excess emission in the $\sim$2.0--4.0~keV energy range and the flux of the fluorescent Fe emission line at 6.4~keV. In addition, the reflection component produces a non-negligible contribution to the soft energy range ($E<$2.0~keV).

\item The analysis of the X-ray spectra allowed us to estimate the abundances of the X-ray-emitting gas for the first time. We demonstrated that the X-ray emission from CH~Cyg is produced by processed material from the red giant component. In particular, the {\it XMM-Newton} RGS high-resolution spectroscopy, which primarily traces the soft X-ray emission, constraints C, N and O abundances to values that agree with photospheric abundance values reported in the literature for the red giant companion in CH\,Cyg. Statistically-acceptable models of the {\it XMM-Newton} MOS data also resulted in N, O, Mg, Si and Fe abundances similar to those of the cool companion in CH~Cyg.

\item We corroborated the variable nature of the X-ray properties of CH\,Cyg. It was previously reported that by 2007 the X-ray flux was decreasing, but the analysis of the {\it Chandra} and {\it XMM-Newton} data presented here shows that it has increased again to levels comparable to those detected by {\it ASCA} 24~yr ago. In addition, the fluorescent, He-like and H-like Fe lines exhibit flux variations. These flux variations are not correlated with $L_\mathrm{X}$, but, might be correlated with the variation from the column density of the highly-absorbed component. Such behavior suggests that the density distribution of the accretion disk is highly variable.  

\end{itemize}

The inclusion of an AGN-like reflection component in the spectral modeling provides a leap towards better understanding the physics behind the production of X-rays from SySt. In addition, this component corroborates that the physics behind accreting sources can be studied across vast mass and accretion scales.

\section*{Acknowledgements}

J.A.T. and O.G.-M. thank Fundaci\'{o}n Marcos Moshinsky (Mexico). J.A.T. thanks support from the UNAM PAPIIT project IA101622. M.K.B. thanks Consejo Nacional de Ciencias y Tecnolog\'{i}a (CONACyT) Mexico for research student grant. O.G.-M. acknowledges financial support by the UNAM PAPIIT project IA109123. LS and MB acknowledge support from UNAM PAPIIT project IN110122. Based on observations obtained with {\it XMM-Newton}, an European Space Agency (ESA) science mission with instruments and contributions directly funded by ESA Member States and National Aeronautics and Space Administration (NASA). This research has made use of data obtained from the Chandra Data Archive and software provided by the {\it Chandra} X-ray Center (CXC) in the application packages {\sc ciao}. This research has made use of software provided by the High Energy Astrophysics Science Archive Research Center (HEASARC), which is a service of the Astrophysics Science Division at NASA/GSFC. This work has made a large use of NASA’s Astrophysics Data System (ADS).

\section*{DATA AVAILABILITY}

The processed data presented and discussed in this article will be shared on reasonable request to the corresponding author. The original data sets can be downloaded from public archives.


\appendix

\section{R Aqr}
\label{sec:app}

\begin{figure}
\begin{center}
  \includegraphics[angle=0,width=\linewidth]{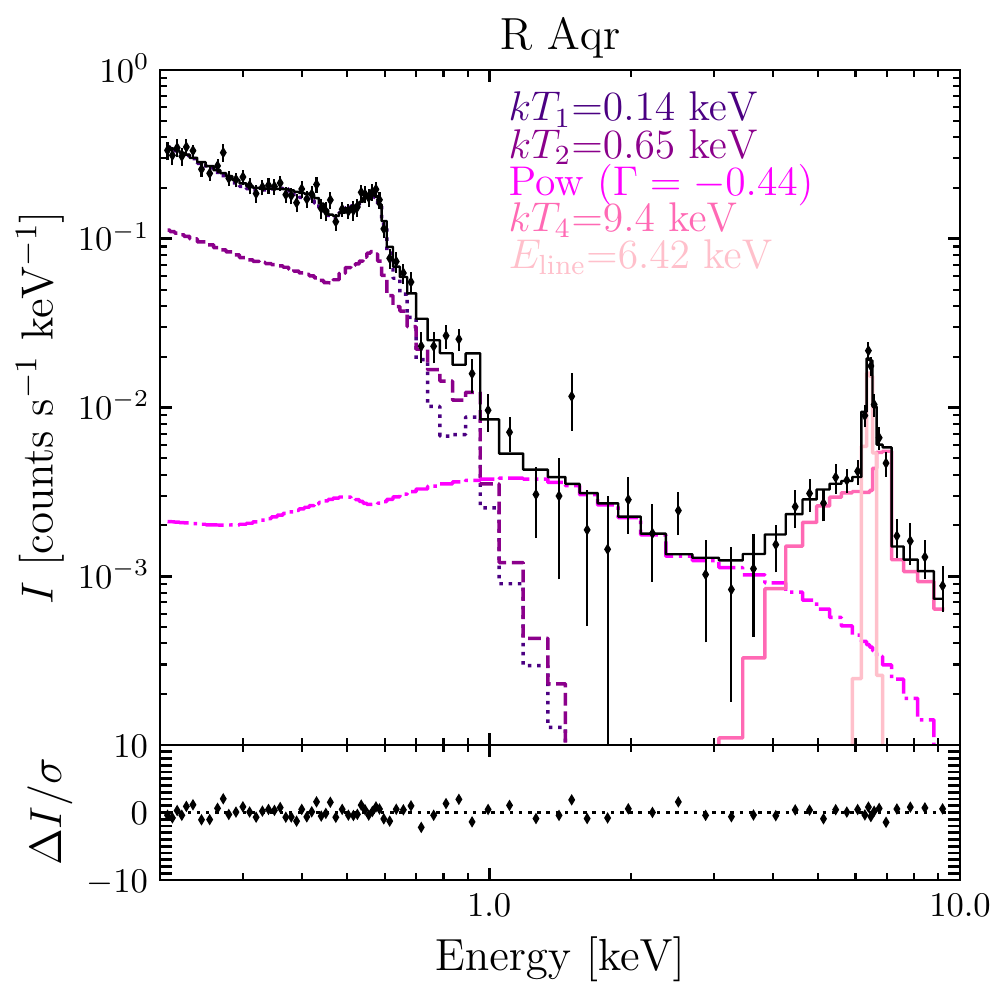}
\caption{Background-subtrated {\it XMM-Newton} EPIC (pn+MOS1+MOS2)
  spectrum of R~Aqr. The black diamonds
  correspond to the observed spectra while the black histograms
  represent the best-fit models. The bottom panel show the
  residuals.}
\label{fig:spec_r_aqr}
\end{center}
\end{figure}

{\it XMM-Newton} EPIC observations of R~Aqr were analysed similarly to those of CH\,Cyg. A combined EPIC (pn+MOS1+MOS2) spectrum was produced and is presented in Fig.~\ref{fig:spec_r_aqr}. This have similar spectral properties as those of CH\,Cyg and other $\beta/\delta$-type X-ray-emitting SySt.

Although the spectrum has a bit worse quality than that of CH\,Cyg, similar models reproduce the spectral properties. In particular, we show in Fig.~\ref{fig:spec_r_aqr} a model including a three-plasma component, a power-law and a spectral line at 6.4~keV. This spectrum is very similar to Model~B of CH\,Cyg listed in Table~\ref{tab:EPIC_par} and presented in Fig.~\ref{fig:spec}. A detailed reflection model as that presented here for CH~Cyg will be pursuit in a subsequent work.

\end{document}